\documentclass[11pt]{llncs}

\usepackage{{amssymb}}
\usepackage{fullpage}
\usepackage[all]{xy}

\newcommand{\remove}[1]{}

\begin{document}

\title{Persistent Asymmetric Password-Based Key Exchange}
\author{Shaoquan Jiang}
\institute{School of Computer Science and Engineering\\ University of Electronic Science and Technology of China\\ \email{Email: shaoquan.jiang@gmail.com}}
\maketitle

\begin{abstract}
Asymmetric password based key exchange is a key exchange protocol
where a client and a server share a low entropic password while  the
server additionally owns a high entropic secret for a public
key. There are simple solutions for this (e.g. Halevi and
Krawczyk (ACM TISSEC 1999) and its improvement by  Boyarsky (CCS 1999)).
In this paper, we consider   a new threat to this type of protocol:  if
a server's high entropic secret gets compromised (e.g., due
to cryptanalysis, virus attack or a poor management), the
adversary might {\em quickly}  break lots  of passwords  and cause uncountable damage.
In this case, one should not expect the protocol to be  secure against
 an off-line dictionary attack since, otherwise, the protocol is
in fact  a secure password-only key exchange where the server also only has a password (by making the server high entropic secret public).
Of course  a password-only key exchange does not suffer from this threat as the  server does not have a high entropic secret at all. However,  known password-only key exchange are not very  efficient (note: we only consider protocols without  random oracles).  This motivates us to  study  efficient and secure asymmetric password key exchange that avoids the new threat.
In this paper, we first provide a formal model for the new threat, where essentially we require that the active adversary can break $\ell$ passwords in $\alpha\ell |{\cal D}|$ steps (for $\alpha<1/2$) only with a probability negligibly close to $\exp(-\beta\ell)$ for some $\beta>0$. Then, we construct a framework of asymmetric password based key exchange.
We prove that our protocol is secure in the usual sense. We also show that it prevents the new threat. To do this, we introduce a new technique by
 abstracting  a probabilistic  experiment from the main proof and providing a neat analysis of it.
\end{abstract}
\section{Introduction}
Key exchange (KE) is one of   the most important issues in secure
communication. It helps two communicants to securely
establish a common session key, with which the subsequent
communication can be protected. In the literature, there are two
types of key exchange. In  type one, two parties
own high entropic secrets (e.g., a signing key of a digital
signature). This type has been extensively studied in the
literature;  see a very partial list
\cite{bck98,K03,ck01,HOW92}. Type two  is  password
authenticated key exchange, in which  it is assumed that the two
parties share a human-memorable (low entropy) password. The major
threat for this type of key exchange is an off-line dictionary
attack. In this case, an adversary  can catch a function value of the
password (say, $F(pw)$). Since the password space is small, he can find the matching password through an exhaustive search. See \cite{Bao03} for an example.
 In the literature,
two types of password key exchange protocols are studied. In the
first type, two parties only own a common password. This type is studied extensively in the literature. In the second type,  the client and server
share a password while the server additionally owns a high entropic
private key of a public key. In this type, there are simple solutions \cite{HK99,Boy99}. In this paper, we consider a new threat to this type of protocols:
 when the server high entropic secret is compromised, the attacker might
quickly break lots of passwords  and cause uncountable
damage. It is desired that the pace he breaks passwords is very slow. Under this,  the server management will have  enough time to realize and defend the attack. Unfortunately, previous  protocols (e.g., \cite{HK99,Boy99}) is not secure against  this threat.

\subsection{Related Work}

The server key leakage problem does not occur in the password-only key exchange protocol since in this setting
 the server does not own a high entropic secret key at all. Hence, an asymmetric password key exchange against this threat is meaningful only if we have a construction that is more efficient than the known password-only protocols. Password-only key exchange was first studied by Bellovin
and Merritt \cite{BM92} and further studied in
\cite{BM93,Jablon97,Lucks97}. The first provably secure solution is
due to Bellare et al. \cite{BPR00} but security holds in the random oracle model which is not our main focus. The first key exchange without random oracles are due to
Goldreich and Lindell \cite{GL01}. But it is very inefficient. The first reasonably efficient solution without random oracles  is KOY protocol \cite{KOY01}  which has 15 exponentiations for each party. This protocol was abstracted into a framework by \cite{GL03} and improved by Gennaro \cite{Gen08}(the contribution of the latter is to remove the signature).  Jiang and Gong \cite{SAC04} (recently abstracted into a framework by \cite{GK10}) constructed an efficient protocol, where using the fastest CCA2 secure encryption \cite{HK09} it costs  5 exponentiations for a client and 6 exponentiations for a server. Katz and   Vaikuntanathan \cite{KV10} constructed a one-round password-only key exchange but less efficient than \cite{SAC04,GK10}.

Asymmetric password based technique was initiated  by Gong \cite{Gong93}.
Halevi and Krawczyk \cite{HK98} (also full version \cite{HK99}) proposed a very efficient asymmetric password based key exchange,
which essentially  let the client use a CCA2 secure encryption to encrypt the password information.
Using encryption \cite{HK09}, this protocol only needs  about two exponentiations for the client and one exponentiation for the server.
It was later  improved by Boyarsky
\cite{Boy99} for security   in the multi-user setting.  However, neither of two protocols
 can  prevent the new threat above because  the  password is encrypted under a server public key and can be adversely decrypted without a dictionary attack if the  private key is leaked.

\subsection{Contribution}

We first provide a formal model for the above server key leakage  problem. It essentially requires  that an adversary can break $\ell$ passwords in
$\alpha\ell |{\cal D}|$ steps (for $\alpha<1/2$) only with probability negligibly close to $\exp(-\beta\ell)$ for some $\beta>0$. Under this assertion, the adversary can not quickly break a lots of passwords. Then, we construct a framework of asymmetric password based key exchange. Our construction is based on a tag-based projective hash family that is modified from projective hash family (tag-PHF) of Cramer-Shoup. We show that our framework is secure in the multi-user setting of  \cite{Boy99} (under  a different formalization, where our approach is a new quantification on the  authentication failure). Our proof does not rely on the  random oracles.  We also prove that our framework is persistent, where we introduce a new technique to achieve this, which is a probabilistic experiment extracted from the main proof. We provide a neat analysis for this experiment.  Our persistency holds in the random oracle model. It is open to construct a protocol whose  security and persistency both hold without random oracles.   We instantiate our framework with a concrete tag-PHF. Our realization only costs 4 exponentiations for the client and 2 exponentiations for the server, which is significantly more efficient than all the known password-only key exchange.

\vspace{0.10in}\noindent{\bf Notions.}\quad    For a set $S$, $x
\leftarrow S$ samples $x$ from $S$ randomly; $A|B$ means
concatenating  $A$ with $B$. We use $negl: \mathbb{N} \rightarrow
\mathbb{R}$ to denote a {\bf negligible} function:  for any
polynomial $p(x)$, $\lim_{n\rightarrow \infty} negl(n)p(n)=0.$
Probability distance of two random variables $A, B$ over set
$\Omega$ is defined as $$\textsf{dist}[A, B]=\frac{1}{2}\sum_{v\in
\Omega}|\Pr[A=v]-\Pr[B=v]|.$$ For $a\in \mathbb{N}$, define $[a]=\{1,
\cdots, a\}$.  PPT means probabilistic polynomial time.

\section{Security Model}

\vspace{0.05in} In this section, we introduce a  security model for asymmetric password key exchange, which is slightly modified from the password-only setting of  Bellare, {\em et al.} \cite{BPR00}. Before proceeding, we first give some notions.
\begin{itemize}
\item[$-$] ${\cal D}$:\hspace{.10in} a password dictionary. For simplicity, assume ${\cal D}=\{1, \cdots, N\}$ with a uniform distribution. But
 our result holds without the uniformity.
\item[$-$] Client $C_i$ and Server $S$: \ \ Asymmetric password key exchange runs between  a client $C_i$ and a server $S$. $S$
has a public key $\Theta$ and a private key $\theta$. He also
shares a password $\pi_i$ with  $C_i$. $\Theta$ is
known to all clients.

\item[$-$] $\Pi_i^{\ell  _i}$ and $\Pi_S^{\ell  _S}$:  \ \ $\Pi_i^{\ell  _i}$ is a protocol instance
$l_i$ within client $C_i$, where  $\ell  _i$ is unique
within $C_i$ for  distinguishing different
instances in $C_i$ but it is not necessarily globally unique.
Similarly, $\Pi_S^{\ell  _S}$ is a protocol instance $\ell  _S$ within Server $S$.
In this paper,
by a general $\Pi_U^{\ell  _U}$, we mean $U$ is  either $S$ or
some client $i.$
\item[$-$] ${Flow}_i$: \ \ The $i$th message in the protocol execution.

\item[$-$] $\textsf{sid}_U^{\ell  _U}$: \ \
session identifier of $\Pi_U^{\ell  _U}$, where $U$ is either a client $i$ or server $S$.
This variable is defined for
security analysis only. Essentially, if  two instances are jointly
executing a protocol then they have the same \textsf{sid}. \textsf{sid} is
clear only when the protocol description is available.

\item[-] $sk_U^{\ell  _U}$:\quad  session key defined by  instance $\Pi_U^{\ell  _U}.$

\item[$-$] $\textsf{pid}_U^{\ell  _U}$: \ \ the party
$\Pi_U^{\ell  _U}$ presumably interacts  with.

\item[-] \textsf{stat}$_U^{\ell  _U}$: \quad session state of $\Pi_U^{\ell  _U}.$ Simply,
it is the intermediate
 data (other than the long term secret) necessary for  the remaining execution of $\Pi_U^{\ell  _U}.$
 If $\Pi_U^{\ell  _U}$ finishes successfully,
  by default   \textsf{stat}$_U^{\ell  _U}=(U, \textsf{pid}_U^{\ell  _U}, sk_U^{\ell  _U}).$

\item[-] \textsf{Client}$(\Pi_U^{\ell  _U}):$ \quad For any $\Pi_U^{\ell  _U}$, either $U$ or $\textsf{pid}_U^{\ell  _U}$ (but not both) is some client.
Hence, it is well-defined if we  use
$\textsf{Client}(\Pi_U^{\ell  _U})$ to denote this client.

\end{itemize}
{\bf Partnering.} \quad $\Pi_U^{\ell  _U}$ and
$\Pi_V^{\ell  _V}$ are {\em partnered} if (1) $\textsf{pid}_U^{\ell  _U}=V$ and
$\textsf{pid}_V^{\ell  _V}=U$; (2) $\textsf{sid}_U^{\ell  _U}=\textsf{sid}_V^{\ell  _V}$.

\vspace{0.05in} \noindent{\bf Adversarial Model.}  \hspace{0.05in}  There are $n$ clients $C_1, \cdots, C_n$ and a Server $S.$ A client $C_i$ will be initialized with
a random password $\pi_i\in {\cal D}$, which is shared with his server $S$. Server $S$,
besides owning all clients' passwords, additionally has a high entropic
public key $\Theta$ and a private key $\theta$. $\Theta$ is also available to all clients.
An adversary can fully control
the
network. He can inject, modify, block messages.
He can also request any session key. Formally, his behaviors are modeled as
access to the following oracles.

\vspace{0.03in}\noindent ${\bf Execute}(i, \ell  _i, S, \ell  _S)$.\quad When this oracle is
called,  a protocol execution between $\Pi_i^{\ell  _i}$ and
$\Pi_S^{\ell  _S}$ takes place. Finally, a
complete message transcript
is returned. This oracle call models an eavesdropping
attack. Note, literately, it can be replaced by a sequence of
\textsf{Send} queries blow.  But it is defined separately by  requiring  that
 \textsf{Execute} queries should not increase  adversary success probability.

\vspace{0.03in}\noindent ${\bf Send}(d, U, \ell  _U, M).$\quad  Upon this query, $M$ is sent to $\Pi_U^{\ell  _U}$ as
$Flow_d$.  The
 output is whatever $\Pi_U^{\ell  _U}$ returns.
By default, when $d=0$, $M=null$.
 This query  models
active attacks.

\vspace{0.03in}\noindent \textbf{Reveal}$(U, \ell  _U).$ \hspace{0.05in}
 When this oracle is called, session key $sk_U^{\ell  _U}$ (if any) is returned. it models a
 session key loss attack.

\vspace{0.03in}\noindent ${\bf Corrupt}(i).$ \quad
Upon this query, $C_i$'s password $\pi_i$ as well as his session states
$\{\textsf{stat}_i^{\ell  _i}\}_{\ell  _i}$  is given
to
 adversary. After this,
 his role  will be taken by adversary. This query  models a break-in
attack or insider attack. We assume $S$ is never corrupted (although a weak corruption of $S$ will be considered when defining persistency in the next subsection).

\vspace{0.03in}\noindent ${\bf Test}(U, \ell  _U). $\quad This query is a security test for session key $sk_U^{\ell  _U}$.
The adversary is allowed to query it only once. The queried session
must have  successfully completed. Throughout the game,  $U$ and
$\textsf{pid}_U^{\ell  _U}$ should not be corrupted; $\Pi_U^{\ell  _U}$
and its partnered instance (if any) should not be issued a {\bf
Reveal} query. When {\bf Test} oracle is called, it flips a fair
coin $b.$ If $b=1,$ then $sk_U^{\ell  _U}$ is
provided to adversary; otherwise, a random number of the same length
is provided.  The adversary then tries to output a guess bit $b'$.
If $b'=b,$ he will be informed {\bf Success}; otherwise, {\bf Fail}.

\vspace{0.05in} We now define the protocol
security, which considers three properties:  correctness, authentication
and secrecy.

\vspace{0.05in} \noindent{\bf Correctness.} \quad
If two partnered instances
accept, they derive   the same   session key except for negligible probability.

\vspace{0.05in} \noindent {\bf Authentication}.\quad  If
some $\Pi_U^{\ell  _U}$, with $U$ and $\textsf{pid}_U^{\ell  _U}$
uncorrupted, has successfully completed
 while it does not have  a {\em unique} partnered instance, then we say
authentication is {\em broken}, denoted by event {\bf Non-Auth}.  Note that  since the password dictionary ${\cal D}$ is small, one can always break
the authentication by guessing a client's password and impersonating him to $S$ (through {\bf Send} queries). Hence, if an adversary makes at most $Q$ {\bf Send} queries, we can only hope that $\Pr[\mbox{\bf Non-Auth}]=Q/|{\cal D}|+negl(\kappa).$ However, this requirement  is not enough.  Boyarsky \cite{Boy99}  discussed an authentication problem  against \cite{HK98} which does not violate this requirement. Intuitively, in his attack, an adversary first obtains a transcript $tr$
 between $C_i$ and $S$; then he corrupts $C_j$ and
obtains $\pi_j$; next he, in the name of $C_j$, communicates with
$S$ under the help of $tr$. The last stage is launched many times
and finally it can obtain $\pi_i$ and hence can impersonate $C_i$
successfully.  The significance of this attack is that a malicious $C_j$ can break another user's password just through repeated attempts to login his own account. In this case, the rule that $N$ consecutive failures  of login results in his account closure   can be easily defeated during his attack, by $N-1$ malicious login attempts followed by one correct login.
We remark that this attack does not occur in a password-only key exchange essentially  because the server only has a password and hence  when $C_j$ attempts to key exchange with $S$ in his own name, the server's answer  can be computed  by himself. That is, an   interaction with $S$ in his own name is useless.
To address the above attack, we  consider the authentication between Client $C_i$ and Server $S$ for each $i$ individually.  Define
 {\bf Non-Auth}$_i$ to be the event {\bf Non-Auth} such that the client in this event is $C_i$. Obviously,  {\bf Non-Auth}$_i, i=1, \cdots, n$ are mutually disjoint and $\vee_{i=1}^n$ {\bf Non-Auth}$_i$={\bf Non-Auth}.
{\em  Our authentication property is to require that for each $i$,
 $\Pr[\mbox{\bf Non-Auth}_i]\le Q_i/|{\cal D}|+negl(\kappa),$} where $Q_i$ is the number of {\bf Send}$(d, U, \ell_U, \cdot)$ queries such that $\textsf{Client}(\Pi_U^{\ell_U})=i.$
Under our definition, interactions between $C_j$ and
$S$  are not counted into $Q_i$ and hence can not increase the probability to break $\pi_i$.

\vspace{0.05in} \noindent{\bf Secrecy}. \quad An
adversary can succeed in a {\bf Test} session. Denote this event by  $\textsf{Succ}.$
Since {\bf Non-Auth} already implies a break of  the protocol, we only consider $\textsf{Succ}$ under  $\neg${\bf Non-Auth}. As an  adversary has a naive success of probability 1/2, we require
$\Pr[{\bf Succ}({\cal A})|\neg\mbox{\bf Non-Auth}]<1/2+negl(\kappa)$.

\vspace{0.05in} Note it is crucial to properly define
session id \textsf{sid} (hence {\em partnership}) so that we do
not classify secure protocols as insecure. For instance (see \cite{SAC04}), if we
define a complete protocol transcript as a session id, then
 any protocol  is insecure since  as long as we hold the last message, {\bf Non-Auth} occurs.
 More subtleties of defining  \textsf{sid} and partnership can be seen in \cite{CBHM04}.  Now we are ready to state the security definition.
\begin{definition} \label{def: sec} Let $Q_i$ be $\sharp$ of \textsf{\em Send}$(d, U, \ell  _U, \cdot)$ queries such that
$i=\textsf{\em Client}(\Pi_U^{\ell  _U}).$  Then, an asymmetric  password key exchange protocol
is {\bf secure} if

\vspace{0.05in} \noindent\textbullet\quad  {\bf Correctness.}

 \noindent\textbullet\quad {\bf Authentication.}
$\Pr[\mbox{\bf Non-Auth}_i]\le \frac{Q_i}{|{\cal D}|}+negl(\kappa), \forall i$.

 \noindent\textbullet\quad {\bf Secrecy. }
$\Pr[{\bf Succ}({\mathcal A})\mid \neg\mbox{\bf
Non-Auth}]<1/2+{negl}(\kappa).$
  \label{def: sec}
\end{definition}

Note if $Q$ is
$\sharp$ of {\bf Send} queries, then $Q=\sum_{i=1}^n Q_i. $ Hence,
authentication in Definition \ref{def: sec} implies that
$\Pr[\mbox{\bf Non-Auth}]<\frac{Q}{|{\cal
D}|}+negl(\kappa).$ This futher indicates that $\Pr[{\bf
Succ}({\mathcal A})]<1/2+\frac{Q_s}{2|{\cal D}|}+{negl}(\kappa),$
which is the security  definition \cite{BPR00} for  the  password-only key exchange.

\subsection{Persistency against Server Key Leakage}

We now formalize the security when the server high entropic  key gets
compromised.  This threat is possible due to  cryptanalysis,  virus attack or  a poor management. In this case, we can not hope the protocol is secure against an off-line dictionary attack as otherwise
the protocol is in fact a secure password-only protocol (by making the server secret public).
We thus consider a weaker guarantee:   the adversary should  not be able to quickly break lots of passwords. Under this assertion, the manager will have enough time
to realize and defend the attack. We remark that previous protocols
\cite{HK98,HK99,Boy99} do not prevent this threat since they
essentially encrypt a password using the servery pub key.

It is desired that  if an attacker intends to break $\ell$ passwords, he has to do so using an dictionary attack individually on each password and in average costs $\ell|{\cal D}|/2$ dictionary guesses.  Quantitatively, if any adversary runs $T<\alpha\ell|{\cal D}|$ steps, then he
can break $\ell$ passwords with probability at most $\exp(-\beta\ell)+negl(\kappa)$ for some $\beta>0$,  where one step is essentially the cost of one dictionary guess and will be defined when  the protocol description is available.  Also note that since $\ell$ does not
necessarily depend on the security parameter $\kappa,$ we can not simply require  the above adversarial success probability be $negl(\kappa)$.
We notice that  it is hard to  tell whether an adversary  has broken a password $\pi_i$ or not. Hence, we can not directly use this definition. However, if this occurs,
it should be easy for him to successfully impersonate client $i$, in which case \textsf{Non-Auth}$_i$ occurs.
Hence, we instead define the adversary success as  the occurrence of \textsf{Non-Auth}$_i$ for at least $\ell$ different $i.$
Finally, we define  the adversary capability. Since persistency only considers a attack that occurs under a very rare circumstance and continues   in a short time,   oracle queries  other than {\bf Send} are immaterial. We thus formally define the persistency as follows.

\begin{definition} $\ell  \in\mathbb{N}$ and $\alpha<1/2.$  $\Xi$ is an asymmetric password-based key
exchange protocol, where ${\cal D}$ is the password dictionary and
$(\Theta, \theta)$ is the server's public key and private key
pair. Then $\Xi$ is {\bf persistent} if for any PPT adversary ${\cal A}$ that runs  $T<\ell \alpha
|{\cal D}|$ steps with access to {\bf Send} oracles, {\bf
Non-Auth}$_i$ occurs to $\ell$  different $i$ with probability at most  $\exp(-\beta \ell)+negl(\kappa)$ for some $\beta>0$, where a basic step is specified in a concrete protocol.
\end{definition}

 \noindent \section{Tag-Based Hash Proof System}

In this section, we introduce a {\em tag-based hash proof system},
revised from the original hash proof system \cite{CS01} (in fact the brief introduction in \cite{GL03} suffices) by
 adding a tag. Special forms of hash proof system are used by \cite{GK10,KV09,KV10,GL03,Gen08} to construct password-only key exchange protocols.

\subsection{Subset Membership Problem }
A hard
subset membership problem essentially is a problem that one can
efficiently sample a hard instance in it. Formally,
\textsf{a subset membership problem ${\cal I}$} is a collection
$\{{\cal I}_n\}_{n\in \mathbb{N}},$ where ${\cal I}_n$ is a
distribution for a random variable $\Lambda_n$ that  can be sampled in polynomial time:
\begin{itemize}
\item[\textbullet] Generate a {\em finite} non-empty set $X_n, L_n\subseteq \{0,
1\}^{poly(n)}$ s.t. $L_n\subset X_n$, and  distribution $D(L_n)$
over $L_n$ and  distribution $D(X_n\backslash L_n)$ over
$X_n\backslash L_n.$

\item[\textbullet] Generate a witness set $W_n\subseteq \{0,
1\}^{poly(n)}$ and a {\bf NP}-relation $R_n\subseteq X_n\times W_n$ such
that $x\in L_n$ if and only if there exists $w\in W_n$ s.t. $(x,
w)\in R_n.$ $x\leftarrow D(L_n)$ can be sampled in polynomial time and the sampling procedure also  outputs
a witness $w\in W_n$ s.t.
$(x, w)\in W_n.$ We use $x\stackrel{w}{\leftarrow} D(L_n)$ to denote
this procedure.  When $w$ is not a concern, we omit it. Further,
$x\leftarrow D(X_n\backslash L_n)$ can be also sampled in polynomial time.
\end{itemize}
Finally denote $\Lambda_n=\langle X_n, L_n, W_n, R_n, D(L_n),
D(X_n\backslash L_n)\rangle.$ ${\cal I}=\{{\cal I}_n\}_{n\in
\mathbb{N}}$ is called a {\bf hard subset membership problem} if for
$\langle X_n, L_n, W_n, R_n, D(L_n), D(X_n\backslash
L_n)\rangle\leftarrow {\cal I}_n$,
 $x$ and $y$ are
indistinguishable when $y\leftarrow D(X_n\backslash L_n), x\leftarrow D(L_n)$.

\subsection{Tag-based Projective Hash Function}
Let $\Lambda=\langle X, L, W, R, D(L),
D(X\backslash L)\rangle$ be sampled from a hard subset membership problem
${\cal I}_n$. Consider a tuple $\Psi=\langle {\cal H}, {\cal
K}, X, L, G, S, \alpha\rangle $, where  $G, S,
{\cal K}$ are finite but non-empty sets, ${\cal H}=\{H_{k}(\cdot, \cdot)\mid k\in {\cal
K}\}$ is  a set of functions from $X\times \{0, 1\}^*$ to $G$ and $\alpha: {\cal
K}\rightarrow S$ is a deterministic function.  ${\cal K}$ is called
a {\em key space}, $k\in {\cal K}$ is called the {\em projection
key}; $S$ is called the {\em projection space} for $\alpha$.
$\Psi $ is called a
\textsf{tag-based projective hash function} (tag-PHF) for $\Lambda$ if for any $x\in L$ and tag $z\in\{0, 1\}^*$, $H_{k}(z, x)$  is uniquely determined by $\alpha(k), z, x$. It is
called an \textsf{efficient tag-PHF} if $\alpha(k)$ and $H_{k}(z, x)$ are
both polynomially computable from  $(k, x, z)$ and if $H_{k}(z, x)$ also is
polynomially computable from  $x, w, \alpha(k), z$ where $(x, w)\in
R$. In this paper, by \textsf{tag-PHF}, we mean an efficient \textsf{tag-PHF}.

The following notion of {\em computational universal$_2$} is
slightly revised from \cite{HK07}, which in turn is extended from
the notion of  {\em universal$_2$} by relaxing the statistical
indistinguishability to the computational indistinguishability.

\begin{definition}  $\Lambda=\langle X, L, W, R, D(L), D(X\backslash
L)\rangle\leftarrow {\cal I}_n$, where $\{I_n\}_n$ is a hard subset
membership problem. $\Psi=\langle {\cal H}, {\cal K}, X, L, G, S,
\alpha\rangle$ is a tag-based projective hash function for
$\Lambda.$ $\Psi$ is {\bf computational universal$_2$} if any PPT
${\cal A}$ only has a negligible advantage in the following game.
Take   $k\leftarrow {\cal K}$ and provide $(\Psi, \alpha(k))$
to ${\cal A}$. ${\cal A}$ can do the following.
\begin{itemize}
\item[-] ${\cal A}$ can adaptively query   $(z, x)\in\{0, 1\}^*\times X$ to an {\bf Evalu} oracle, where
oracle {\bf Evalu} is defined as follows.   It checks if $x\in L$
(maybe in exponential time). If yes, return $H_k(z, x)$; otherwise
$\perp$.
\item[-]  ${\cal A}$ can ask {\em once} to compute  some $(z_1, x_1)\in \{0, 1\}^*\times X\backslash L$.
 In turn, he will receive $H_k(z_1, x_1)$.

\item[-] ${\cal A}$ can ask  {\em once} to test some  $(z_2, x_2)\in  \{0, 1\}^*\times X\backslash L$ for $(z_2, x_2)\ne (z_1, x_1).$ In turn, he will
receive $K_b$, where $b\leftarrow \{0, 1\}, K_0=H_k(z_2, x_2)$ and
$K_1\leftarrow G.$
\end{itemize}
Finally, ${\cal A}$ outputs bit $b'$ for guessing $b$ and
succeeds if $b'=b.$
 \label{def: phf}
\end{definition}

\subsection{A Useful Lemma}

\vspace{0.05in} \noindent   $\{{\cal I}_\kappa\}_{\kappa}$ is  a
hard subset membership problem. Take $\Lambda=\langle X, L, W, R, D(L), D(X\backslash L)\rangle \leftarrow {\cal I}_\kappa$.
Define a tag-based PHF $\Psi=\langle {\cal H}, {\cal K}, X, L, G, S, \alpha\rangle$
 for $\Lambda$, where  $G=\{0,
1\}^{2\kappa}.$ Take $k\leftarrow {\cal K}$ as a private key,
 $pk=(\alpha(k)$, $desc(\Psi))$ as a public key where $desc(\Psi)$ is the description of $\Psi$.
Let $\textsf{MAC}: \{0, 1\}^*\rightarrow \{0,
1\}^\kappa$ be a message authentication code with key space $\{0, 1\}^\kappa.$ Consider the following game between a PPT adversary ${\cal A}$ and a challenger, where
${\cal A}$ receives $pk$ and challenger
keeps  $k$.  Let $\Theta=\{ \}$
and $c\leftarrow \{0, 1\}$.
\begin{itemize}
\item[\textbullet] {\bf Challenge Query. } \hspace{0.10in} ${\cal A}$ can adaptively
query with any  tag $z$. Upon this, challenger takes
$x\stackrel{w}{\leftarrow} L$, lets $(a_{0}, s_{0})=H_k(z, x)$,
$(a_{1}, s_{1})\leftarrow \{0, 1\}^{2\kappa}$,  returns $(x, a_{c}, s_c)$
and updates $\Theta=\Theta\cup \{(z, x, a_c, s_c)\}.$

\item[\textbullet] {\bf Compute Query. } \hspace{0.10in} ${\cal A}$ can adaptively query
 with $(z, x, \sigma, m)$. If $(z, x, a', s')\in
\Theta$ for some $a', s'$, let $a=a', s=s'$; otherwise, let $(a,
s)=H_{k}(z, x)$. If $\sigma= \textsf{MAC}_a(m)$, return $(a, s)$; otherwise $\perp$.
\end{itemize}
At the end of the game, ${\cal A}$ outputs a guess bit $c'$ for $c.$ He succeeds if $c'=c.$

Denote this game by $\Re$. The lemma below states  that ${\cal A}$ only has a negligible advantage (see Appendix A for proof).

\begin{lemma}  $\{{\cal I}_\kappa\}_\kappa$ is a hard subset membership problem, $\Psi$ is
computational universal$_2$ and $\textsf{\em MAC}$ is existentially unforgeable. Then
$\Pr[{\bf Succ}({\cal A})]=1/2+negl(\kappa)$.  \label{le: CU2}
\end{lemma}

\section{Red Ball Experiment}

We consider an experiment:
there are $n$ boxes, where each box contains $a$ identical balls except for a color difference,  where one of them is colored red while the remaining $a-1$ balls are
colored white. Algorithm ${\cal A}$ adaptively draws $t$ balls from these boxes. Each
time it chooses a box and then draws a ball uniformly randomly from it
without replacement. Let $\ell  \in \{1, \cdots, n\}$.
 We use $\Theta^{\cal A}_{t,
n, \ell  }(a_1, \cdots, a_n)$ to denote the success probability that
algorithm ${\cal A}$ draws $t$ balls (from these boxes) such that  $\ell  $ of
them are red, where box $i$ initially contains  $a_i$ balls including one red. When the red ball in the box is taken, set
$a_i$=0 since ${\cal A}$ knows all are white in this box and does
not need to draw any ball from it any more. Let $\Theta_{t, n, \ell
}(a_1, \cdots, a_n)=\max_{\cal A} \Theta^{\cal A}_{t, n, \ell }(a_1,
\cdots, a_n)$. It is easy to see that for any permutation $(a_1',
\cdots, a_n')$ of $(a_1, \cdots, a_n)$, $\Theta_{t, n, \ell }(a_1,
\cdots, a_n)=\Theta_{t, n, \ell  }(a_1', \cdots, a_n')$ holds.  We prove the following important lemma, where the proof is by induction. Due to the page limit, the details are  in Appendix B.

\begin{lemma} If $1\le a_1\le a_2\le\ldots\le a_n, 0\le \ell\le n, t\ge 0,$  then
\begin{equation}\Theta_{t, n, \ell  }(a_1, \cdots,  a_n)=\Pr\Big{[}\sum_{i=1}^\ell x_i\le t:
\quad x_i\leftarrow [a_i]\Big{]}.
\label{eq: reduce}
\end{equation}  \label{le: reduce}
\end{lemma}

\begin{figure*}[!ht]
$$
\input xy
\xyoption{all}  \xymatrix@R=0.05in@C=1.8in{
C_i(\pi_i)&S(\pi_{i},\theta) \\
\txt<12pc>{$x\stackrel{w}{\leftarrow} D(L),\ y=\textsf{T}(\pi_i, x)$\\ $(k_0, k_1)=H_\theta(i, x)$ \textsf{using} $w$\\
$\tau_0=\textsf{MAC}_{k_0}(C_i|S|y)$} \ar[r]^-{C_i\ |\ y\ |\ \tau_0 } &*\txt<12.3pc>{$x=\textsf{T}^*(\pi_i, y),$
$\zeta\leftarrow \{0, 1\}^\kappa$\\
$(k_0', k_1')=H_\theta(i, x)$ \textsf{using} $\theta$\\
$\tau_0\stackrel{?}{=}\textsf{MAC}_{k_0'}(C_i|S|y)$\\
$\omega=C_i|S|y|\zeta$,\ $\tau_1=\textsf{MAC}_{k_0'}(\omega|1)$} \\
\txt<12.6pc>{
$\omega=C_i|S|y|\zeta$,\ $\tau_1\stackrel{?}{=}\textsf{MAC}_{k_0}(\omega|1)$\\
$\tau_2=\textsf{MAC}_{k_0}(\omega|2)$, \textsf{output } $sk=k_1$}
 &\ar[l]_-{S\ |\ \tau_1\ |\ \zeta}
 \txt<12pc>{} \\
\txt<12pc>{}\ar[r]^{\tau_2} &
\txt<12pc>{$\tau_2\stackrel{?}{=}\textsf{MAC}_{k_0'}(\omega|2)$,
$\textsf{output } sk=k_1'$}}
$$
\begin{center}
\caption{Password Key Exchange Framework \textsf{HPS-PAKE} (details in the bodytext)}
 \label{fig: frame}
\end{center}
\end{figure*}
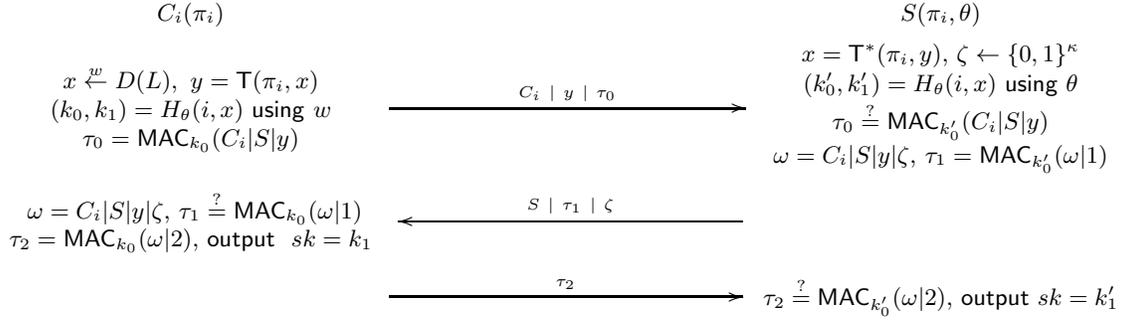

\begin{theorem} If $t<\alpha\ell a$ and $\alpha<0.5$, then

\hspace{0.5in} $\Theta_{t, n, \ell }(a, \cdots, a)<\exp({-2(0.5-\alpha)^2\ell}).$
\label{thm: prob}
\end{theorem}
\noindent{\bf Proof. } By Lemma \ref{le: reduce}, $\Theta_{t, n,
\ell }(a, \cdots, a)$ equals
 $$\begin{array}{lll}
&& \Pr[x_1+\cdots+x_{\ell }\le t]\\
 &{=}&\Pr[\frac{\sum_{i=1}^{\ell
}x_i}{\ell }-\frac{a}{2}\le
-(\frac{a}{2}-\frac{t}{\ell })]\\
&\stackrel{*}{\le}& \exp({-2\delta^2\ell /a^2}), \hspace{0.20in}
\delta=\frac{a}{2}-\frac{t}{\ell }>(0.5-\alpha)a\\
&\le& \exp({-2(0.5-\alpha)^2\ell}),
\end{array}
 $$
where inequality $(*)$ holds since ${\bf E}[x_i]=\frac{a}{2}$ and
the Hoefding inequality. $\hfill\blacksquare$

\section{Our PAKE Framework}
We now introduce our client-server password key exchange framework.
Let ${\cal I}=\{{\cal I}_\kappa\}_\kappa$ be a hard subset
membership problem and $\Lambda=(X, L, W, R, D(L), D(X\backslash
L))\leftarrow {\cal I}_\kappa$.  $\Psi=({\cal H}, {\cal K}, X, L, G, S, \alpha)$ is a tag-based projective hash
family  for
$\Lambda$, where $G=\{0, 1\}^{2\kappa}$. ${\cal D}=\{1, \cdots, N\}$
is the set of all possible passwords with uniform distribution.  We say $\textsf{T}, \textsf{T}^*:
{\cal D}\times X\rightarrow X$ are  a
 \textsf{regular transformation pair}  if they are efficiently computable and also satisfy the following.

\vspace{0.05in}\noindent {\bf R-1}.\quad For  any fixed $\pi\in{\cal D}$,
$\textsf{T}^*(\pi, \textsf{T}(\pi, x))=x$,  $\forall x\in X.$ i.e., $\textsf{T}^*(\pi, \cdot)$ is
the inverse function of $\textsf{T}(\pi, \cdot)$.

\vspace{0.05in}\noindent {\bf R-2}. \quad For any $y\in X$, there is at most one $\pi\in {\cal D}$ such that
$\textsf{T}^*(\pi, y)\in L.$

    \vspace{0.05in} $\textsf{MAC}_k:\{0,
1\}^*\rightarrow \{0, 1\}^\kappa $ is a secure message
authentication code. The setup is as follows. For the server $S$,
take $\theta\leftarrow {\cal K}$ and compute
$\Theta=\alpha(\theta).$ $\theta$ will be the private key for $S$
and $\Theta$ will be his public key. $\Theta$ is known to all
clients. For each client $C_i$, take $\pi_i\leftarrow {\cal D}$ as
the password for $C_i$, shared with $S$. $C_i$ stores  $\Theta$
publicly  and $\pi_i$ secretly. $S$ stores $\pi_i, \theta$ secretly
and $\Theta$ publicly. The key exchange protocol between $S$ and
$C_i$ is carried out as follows (also see Fig. \ref{fig: frame}), where  we assume that $y\in X$ has been verified but in Section \ref{sec: concrete} we will remove this condition with almost zero price for a concrete and  efficient realization of our framework.

\begin{itemize}
\item[1.] $C_i$ takes $x\stackrel{w}{\leftarrow} D(L)$. Then he uses $\pi_i$ to
compute $y=\textsf{T}(\pi_i, x)$, computes $(k_0, k_1)=H_\theta(i,
x)$ using $w, x, \Theta$, and generates
$\tau_0=\textsf{MAC}_{k_0}(C_i|S|y).$ Finally, he sends
$C_i|y|\tau_0$ to server $S.$ $C_i$ sets his session state
 $\textsf{stat}=C_i|S|y|k_0|k_1.$

\item[2.] Receiving $C_i|y|\tau_0$, server $S$ uses $\pi_i$ to de-transform $y$ back
to $x=\textsf{T}^*(\pi_i, y),$ computes $(k_0', k_1')=H_\theta(i,
x)$ using $(\theta, x).$ He then verifies if
$\tau_0\stackrel{?}{=}\textsf{MAC}_{k_0'}(C_i|S|y).$ If no, reject;
otherwise, he takes $\zeta\leftarrow \{0, 1\}^\kappa$ and computes
$\tau_1=\textsf{MAC}_{k_0'}(\omega|1)$ for $\omega=C_i|S|y|\zeta$.
Finally, he sends $S|\tau_1|\zeta$ to $C_i.$ $S$ sets his session state
$\textsf{stat}=C_i|S|y|\zeta|k_0'|k_1'.$

\item[3.] Upon $S|\tau_1|\zeta$, $C_i$ verifies if $\tau_1\stackrel{?}{=}\textsf{MAC}_{k_0}(\omega|1)$ for $\omega=C_i|S|y|\zeta$.
If no, reject; otherwise, he computes and sends
$\tau_2=\textsf{MAC}_{k_0}(\omega|2)$ to $S$ and outputs
session key $sk=k_1.$ $C_i$ updates  $\textsf{stat}=C_i|S|sk.$

\item[4.] Upon $\tau_2$, Server $S$ verifies if $\tau_2\stackrel{?}{=}\textsf{MAC}_{k_0'}(\omega|2)$.
If no, reject; otherwise, output session key $sk=k_1'.$ $S$ updates
$\textsf{stat}=C_i|S|sk.$
\end{itemize}

\vspace{0.05in} \noindent{\bf Remark. } \quad We outline how some attacks are prevented in order to better understand our protocol.  (1) {\em
against impersonation attack. } If attacker impersonates $C_i$ to
generate and send $Flow_1=C_i|y|\tau_0$ to $S$, then since he does
not know $\pi_i$ and hence $\textsf{T}^*(\pi_i, y)\in L$ with
probability $1/|{\cal D}|$. When $x:=\textsf{T}^*(\pi_i, y)\not\in
L$, $\tau_0$ will be rejected since $(k_0, k_1)=H_\theta(i, x)$
appears random to the attacker.  \quad (2) {\em
against insider attack (as in \cite{Boy99}). } When a malicious $C_j$ eavesdrops a
transcript $tr=C_i|y|\tau_0|S|\tau_1|\zeta|\tau_2$ between $C_i$ and $S$, then he executes the protocol with $S$ in the name of himself but using
$tr$ as a help. Toward this, he might send
$Flow_1=C_j|y|\tau_0^*$ to $S$ and hope to receive a response from the latter.  $\tau_0^*$ is acceptable  only if
$\tau_0^*=\textsf{MAC}_{k_0^*}(C_j|S|y)$, where
 $(k_0^*, k_1^*):=H_\theta(j, x^*)$ for  $x^*=\textsf{T}^*(\pi_j,
y)$. The only useful information is $\tau_0$ which is computed using $(k_0, k_1):=H_\theta(i, x)$ for  $x=\textsf{T}^*(\pi_i,
y)$.  However, no matter $\pi_j=\pi_i$ or not, we have that  $(i, x^*)\ne (j,
x)$ as $i\ne j$ ({\bf this is the main reason we use tag-HPS instead
of HPS in this paper}). This allows us to claim that $k_0^*$ and $k_0$ are
computationally independent. If $x\not\in L$, this is automatically true
by computational universal$_2$ definition. In our protocol, even if
$x\leftarrow D(L),$ this computational independency still holds; otherwise,
one can simply reduce to break the hardness of $L$. Thus, $S$ will always reject $\tau_0^*$. Since this rejection occurs without considering the value of $\pi_i$, it follows that
the candidate space of $\pi$ in view of adversary does not reduce.
\quad (3) {\em session key secrecy. } The session key
$sk=k_1$ is computed by $(k_0, k_1)=H_\theta(i, x).$ Client $C_i$
can compute this since he knows the witness $w$ of $x\in L$ and
server $S$ can compute this since he knows  $\pi_i$ (for recovering
$x$ from $y$) and $\theta$ for $(k_0, k_1).$ Any outsider can not
compute $(k_0, k_1)$ since given $x$ and $\Theta$, $H_\theta(i, x)$
is indistinguishable from random, which is implied by Lemma \ref{le: CU2}.

\section{Security}

In this section, we prove the security of our protocol. Before this, we define the session id in the protocol as
$\textsf{sid}_U^{\ell _U}=C_i|S|y|\zeta,$ where $U$ is the client $i$ or server $S.$ Since the password
$\pi_i$ for $C_i$ and $S$ and $\theta$ are both
 fixed after the system initiation, $H_\theta(i, x)$ is determined for given $C_i|S|y.$ Hence,
  two partnered parties must have the same session key. It remains to consider the authentication and secrecy,
  which we will prove  using a game-hopping approach.

\begin{theorem}
${\cal I}=\{{\cal I}_\kappa\}_\kappa$ is a hard subset membership problem.
$\textsf{\em MAC}: \{0, 1\}^*\rightarrow \{0, 1\}^\kappa$ is an existentially unforgeable message authentication code.
$\Psi$ is computational universal$_2$ for ${\cal I}.$
$(\textsf{\em T}, \textsf{\em T}^*)$ is a regular transformation pair. Then \textsf{\em HPS-PAKE} is secure. \label{thm: secure}
\end{theorem}

\noindent {\bf Proof. }  We modify the security game (denoted by $\Gamma^{rea}$)
into games $\Gamma_0(=\Gamma^{rea}), \Gamma_1,
\Gamma_2$ such that any adversary view (hence event {\bf
Non-Auth}$_i$ or {\bf Succ} as they are in the adversary view)
between each neighboring pair are negligibly close. For simplicity,
we regard \textsf{Execute} query as a result of
 4 \textsf{Send} queries (i.e.,
\textsf{Send}$(d, \cdot), d=0, 1, 2, 3$) and later will remove its
effect on {\bf
Non-Auth}$_i$ and {\bf Succ} by analyzing these special \textsf{Send}
queries. For simplicity, we assume  the
\textsf{Normal} condition: \quad sampling $x\leftarrow
D(L)$ never repeats the same $x$ (otherwise, we can break the  hardness of ${\cal I}$: given challenge $x$, sample $y\leftarrow D(L)$. Then $x=y$ for $x\leftarrow D(L)$ holds non-negligibly while $x\ne y$ always holds for $x\leftarrow D(X\backslash L)$).

\vspace{0.05in} \noindent{\bf Game $\Gamma_1$. }\hspace{0.05in}
 We modify $\Gamma_0$ to $\Gamma_1$ with the following differences.
$\textsf{Send}(0, i, \ell _i, null)$ oracle defines $(k_0, k_1)\leftarrow \{0, 1\}^{2\kappa}$ (instead of
 $(k_0, k_1)=H_\theta(i, x)$). $\Gamma_1$ maintains a list ${\cal Q}$ of record  $(i, y, k_0, k_1)$.
For  consistency, $\textsf{Send}(1, S, \ell _S, C_i|y|\tau_0)$ is handled as follows. First check
 if  $(i, y, u_0, u_1)\in{\cal Q}$ for some $(u_0, u_1)$. If no, process normally using $\theta$; otherwise,
 define $(k_0', k_1')=(u_0, u_1)$ and proceed normally.

\begin{lemma}$\textsf{\em View}({\cal A}, \Gamma_0)\approx \textsf{\em View}({\cal A}, \Gamma_1).$ \label{le: 01}
\end{lemma}

\noindent {\em Proof. } If the views of ${\cal A}$ are distinguished by ${\cal D}$, we construct adversary ${\cal B}$ to violate
 Lemma \ref{le: CU2}. Upon $desc(\Psi), \Theta=\alpha(\theta)$, ${\cal B}$ simulates $\Gamma_0$ as follows.
Let ${\cal Q}=\{ \}.$

\vspace{0.05in}\noindent{\bf Send}$(0, i, \ell _i, null).$ \hspace{0.05in} Upon this query, ${\cal B}$
issues a {\tt Challenge} query with tag $i$ and
in turn receives $(x, a_c, s_c)$. He defines $(k_0, k_1)=(a_c, s_c)$
and normally finishes the simulation in this query. Finally, he  define
 $\textsf{state}_i^{\ell _i}=C_i|S|y|k_0|k_1$ and update ${\cal Q}={\cal Q}\cup \{(i, y, k_0, k_1)\}$. Note in this case,
 the challenger of ${\cal B}$ will update his list $\Omega=\Omega\cup\{(i, x, k_0, k_1)\}.$

\vspace{0.05in}\noindent{\bf Send}$(1, S, \ell _S, C_i|y|\tau_0).$
\hspace{0.05in} Upon this query, compute $x=\textsf{T}^*(\pi_i, y)$.
Then, he issues \textsf{Compute} query $(i, x, \tau_0, C_i|S|y)$. In
turn, he will receive $(a, s)$. If $(a, s)=\perp$, he rejects;
otherwise, define $(k_0', k_1')=(a, s)$ and finishes the remaining
simulation in this query normally. In the later case, also update
$\textsf{stat}_S^{\ell _S}=C_i|S|y|\zeta|k_0'|k_1'.$ Note if $x$ was
generated in $\textsf{Send}(0, i, \cdot)$, then $(i, x, a_c, s_c)\in
{\Omega}$. In this case, the simulation is consistent with
$\Gamma_c$:\quad if $\tau_0=\textsf{MAC}_{a_c}(C_i|S|y)$, then
\textsf{Compute} oracle returns $(a, s)=(a_c, s_c)$; otherwise, it
returns $(a, s)=\perp$ (and ${\cal B}$ will correctly reject
$\tau_0$). If $x$ is not generated in $\textsf{Send}(0, i, \cdot)$
(note it could be generated by Client $i'\ne i$), then $(i, x, *,
*)\not\in\Omega$ and hence $\tau_0$ will be verified by the
challenger of ${\cal B}$ using $(k_0,k_1)=H_\theta(i, x)$ computed
using $\theta$. In this case, $(a, s)=\perp$ if $\tau_0$ is invalid;
$(a, s)=(k_0, k_1)$ otherwise. Hence, in any case, the simulation in this query
is perfectly consistent  with
$\Gamma_c.$

\vspace{0.05in}\noindent{\bf Send}$(2, i, \ell _i, S|\zeta|\tau_1)$ \hspace{0.05in}
Upon this case, use $\textsf{stat}_i^{\ell _i}$ to simulate normally. Finally,
if $\tau_1$ is accepted, update $\textsf{stat}_i^{\ell _i}=C_i|S|k_1$.

\vspace{0.05in}\noindent{\bf Send}$(3, S, \ell _S, \tau_2)$ \hspace{0.05in} Upon this case,
use $\textsf{stat}_S^{\ell _S}$ to simulate normally. Finally,
if $\tau_2$ is accepted, update $\textsf{stat}_S^{\ell _S}=C_i|S|k_1'$.

\vspace{0.05in}\noindent{\bf Reveal}$(U, \ell _U)$ and {\bf Test}$(U, \ell _U).$
\hspace{0.05in} This occurs only when $\Pi_U^{\ell _U}$ is successfully completed. In this case,
$sk_U^{\ell _U}$ is well defined in $\textsf{stat}_i^{\ell _i}$ above. Hence, the simulation is  normal.

\vspace{0.05in}\noindent{\bf Corrupt}$(i)$ \hspace{0.05in} As seen
above, $\textsf{stat}_i^{\ell _i}$ is well defined and $\pi_i$ is
known. Hence,  the simulation is normal.

 From the description of ${\cal B}$, we can see that when challenge bit $c=0$, the
 simulated game by ${\cal B}$ is $\Gamma_0$; otherwise, it is $\Gamma_1$.
  Hence, the distinguishability between $\Gamma_0$ and $\Gamma_1$ leads to
  violate  Lemma \ref{le: CU2}.  $\hfill\square$

\vspace{0.10in} \noindent {\bf Game $\Gamma_2.$ } \hspace{0.05in} We
modify $\Gamma_1$ to $\Gamma_2$ as follows. In
oracle \textsf{Send}$(0, i, \ell _i, null)$, take $x\leftarrow X$ (instead of
$x\leftarrow L$). Note since $w$ is not used in the simulation of
$\Gamma_1$, no further change is required toward the consistency
with  this modification. By simply reducing to hardness of ${L}$, we
have
\begin{lemma} $\textsf{\em View}({\cal A}, \Gamma_1)\approx \textsf{\em View}({\cal A}, \Gamma_2).$ \label{le: 12}

\end{lemma}

We analyze $\Gamma_2$. Recall that, in \textsf{Send}$(1, S, \ell _S,
C_i|y|\tau_0)$, when $(i, y, *, *)\not\in{\cal Q}$, we
 define $(k_0', k_1')=H_\theta(i, x)$ and verify $\tau_0$ with
$k_0'$. Consider a \textsf{Bad} event in this query:
 $(i, y, *, *)\not\in{\cal Q}$ and  $\textsf{T}^*(\pi_i, y)\not\in L$ but $\tau_0$ is valid. We
 show

\begin{lemma} $\Pr[{\bf Bad}(\Gamma_2)]=negl(\kappa).$ \label{le: bad}

\end{lemma}

\noindent{\em Proof. }  Assume the lemma is not true.  Let {\em an
irregular query} be  a \textsf{Send}$(1, S, \ell _S, C_i|y|\tau_0)$
query where $(i, y, *, *)\not\in{\cal Q}$ and $\textsf{T}^*(\pi_i,
y)\not\in L$.
 Let $\sharp$ of irregular queries be bounded by $\nu.$
Use $\textsf{Bad}_i$ to represent the event: the $i$th irregular
query is the {\em first} \textsf{Bad} event. Note when \textsf{Bad}
occurs, there exists a unique
 \textsf{Bad}$_i$ event.

We now construct an adversary ${\cal A}'$ to break the computational
universal$_2$ property of $\Psi.$ Upon $desc(\Psi), \Theta$, ${\cal
A}'$ takes $t\leftarrow \{1, \cdots, \nu\}$ and initializes $\pi_i$
for each $C_i$ and simulates $\Gamma_2$, except when he needs to use
$\theta$, which is one of the following scenarios (especially note
$(k_0, k_1)$ in $\textsf{Send}(0, \cdot)$ is taken randomly in $\{0,
1\}^{2\kappa}$ without using $\theta$). (1) $S$ is corrupted and
$\theta$ should be given to ${\cal A}$. This will not occur since we
assume $S$ is uncorrupted; (2) in \textsf{Send}$(1, S, \ell _S,
C_i|y|\tau_0)$, ${\cal A}'$ will use $\theta$ to compute $(k_0',
k_1')$ {\bf in case of  $(i, y,
*,
*)\not\in{\cal Q}$}. In this case, ${\cal A}'$ can compute
$x=\textsf{T}^*(\pi_i, y)$ and query his $\textsf{Evalu}$ oracle to
compute $H_\theta(i, x).$ When $x\in L$, he will receive
$H_\theta(i, x)$; when $x\not\in L$, he will receive $\perp.$ For
the former case, he proceeds normally; for the latter case, it is an
irregular query. If this is the $j$th irregular query for $j<t$,
then he rejects $\tau_0$; if it is the $t$th irregular query, he
issues $(i, x)$ as a challenge query, in turn he will receive $(a_c,
s_c)$ for challenge bit $c$. If
$\tau_0=\textsf{MAC}_{a_c}(C_i|S|y)$, he outputs 0; otherwise 1.
First of all, when $c=1$, $a_c$ is independent of the adversary view
prior to the current query, by unforgeability of \textsf{MAC},
$\tau_0=\textsf{MAC}_{a_1}(C_i|S|y)$ holds negligibly only.
 We ignore this tiny probability.
 When $c=0$ and $t$ is correct,  the adversary view
till the current query is identical to his view in $\Gamma_2$. In
this case, validity of $\tau_0$ is a $\textsf{Bad}_t$ event, in
which
 ${\cal A}'$ must output 0.
 Since \textsf{Bad}$_t$ event implies  that  $\tau_0$ is valid and that upon such an event the simulation by ${\cal A}'$
 prior to the $t$th irregular query is identical to $\Gamma_2.$ (Even without considering the output of ${\cal A}'$ in  the case
 $c=0$ with an incorrect  $t$), we always have that
 $|\Pr[{\cal A}^{'{\bf Evalu}(0, \cdot)}=0]-\Pr[{\cal A}^{'{\bf Evalu}(1, \cdot)}=0]|
\ge \Pr[{\bf Bad}_t(\Gamma_2)]-negl(\kappa)\ge
\frac{\Pr[{\bf Bad}(\Gamma_2)]}{\nu}-negl(\kappa),$ non-negligible, contradiction! Here we use the fact that
when $t$ is random and thus $\Pr[{\bf Bad}_t(\Gamma_2)]=\Pr[{\bf Bad}(\Gamma_2)]/\nu.$ $\hfill\square$

\vspace{0.05in} \noindent For simplicity, we now assume that {\bf Bad} event never occurs.

\begin{lemma} If initiator $\Pi_i^{\ell _i^*}$ accepts $Flow_2^*=S|\zeta^*|\tau_1^*$,
it must have  a unique partner
$\Pi_S^{\ell _S^*}$.  \label{le: part1}
\end{lemma}

\noindent{\em Proof. } Recall that
$\textsf{sid}_i^{\ell _i^*}=C_i|S|y^*|\zeta^*$. Since $S$ will not
sample the same $\zeta^*$ twice, except for a negligible probability
(which we ignore), it follows that the number of partnered instance
$\Pi_S^{\ell _S^*}$ for $\Pi_i^{\ell _i^*}$ is at most one. It
suffices to prove the existence of such $\Pi_S^{\ell _S^*}.$ If it
does not exist, we show \textsf{MAC} is forgeable. Assume
$\textsf{stat}_i^{\ell _i^*}$ after sending $Flow_1$ is
$C_i|S|y^*|k_0^*|k_1^*$. Then, reviewing the definitions of oracles
in $\Gamma_2$, {\em besides computing $\textsf{\em MAC}_{k_0^*}()$
function}, $k_0^*$ (and its identical copy $k_0^{*'}$)  will be used
only in the following scenarios {\em before $\Pi_i^{\ell _i^*}$
verifies $Flow_2^*$}:
 $k_0^*$ is revealed due to the corruption of $C_i$ (note $S$ is uncorrupted), which is impossible since a corrupted party is controlled by ${\cal A}$  and
 so
 $\textsf{Send}(2, i, \ell _i^*, Flow_2^*)$ query would  not have occurred). Hence,
 prior to verifying $Flow_2$ by $\Pi_i^{\ell _i^*}$, $\Gamma_2$ uses $k_0^*$ only for evaluating
 $\textsf{MAC}_{k_0^*}().$ To reduce to the unforgeability of $\textsf{MAC}$,
 it suffices to show that prior to verifying $Flow_2^*$ in $\Pi_i^{\ell _i^*}$, the simulator never evaluates and outputs $\textsf{MAC}_{k_0^*}()$ with input
 $C_i|S|y^*|\zeta^*|1.$ Otherwise, since $\tau_0, \tau_1, \tau_2$ have different input formats, this evaluation must
  be done by $S$ in $\textsf{Send}(1, S, \ell _S, \cdot)$ for some $\ell _S,$
which already implies that $\Pi_s^{\ell _s}$ is partnered with $C_i,$ contradiction to our assumption.
Thus, validity of $\tau_1^*$ implies breaking \textsf{MAC}. $\hfill\square$

\begin{lemma} Assume that $\textsf{\em pid}_S^{\ell _S^*}(:=C_i)$ is  uncorrupted.
If $(i, y^*, \cdot, \cdot)\in{\cal Q}$ in
${\bf Send}(1, S, \ell ^*_S, C_i|y^*|\tau_0^*)$ oracle and  $\tau_2^*$ is accepted in
{\bf Send}$(3, S, \ell ^*_S, \tau_2^*)$, then $\Pi_S^{\ell _S^*}$ has a unique partnered $\Pi_i^{\ell
_i^*}$ for some $\ell^*_i$. \label{le: part2}
\end{lemma}

\noindent{\em Proof. } $\sharp$ of such $\Pi_i^{\ell _i^*}$ is at most one, by $\textsf{Normal}$ condition on $x$. It suffices to prove the existence of $\Pi_i^{\ell ^*_i}$. Assume this is not true.
 By assumption, in $\textsf{Send}(1, S, \ell _S^*,
C_i|y^*|\tau_0^*)$, it holds that $(i, y^*, k_0^*, k_1^*)\in{\cal
Q}$ for some $k_0^*, k_1^*$ and it also holds that
$\tau_0^*=\textsf{MAC}_{k_0^*}(C_i|S|y^*)$ (otherwise, $\tau_0^*$ in
$Flow_1$  was rejected and it would be impossible for
$\Pi_S^{\ell _S^*}$ to verify and accept $\tau_2^*$).
 Hence, the fact that $(i, y^*, k_0^*, k_1^*)$ was recorded in ${\cal
Q}$  implies that  $\Pi_i^{\ell _i^*}$ for some $\ell _i^*$ must have
sampled $x=\textsf{T}^*(\pi_i, y^*)$. By \textsf{Normal} condition,
$\Pi_i^{\ell _i^*}$ is the only instance that samples this value.
Since $\Pi_i^{\ell _i^*}$ is not partnered with $\Pi_S^{\ell _S^*},$
 $\Pi_i^{\ell _i^*}$ does not compute $\textsf{MAC}_{k_0^*}()$ with input
$C_i|S|y^*|\zeta^*|2$, where $\zeta^*$ is generated by $\Pi_S^{\ell
_S^*}$. As in the previous lemma, $k_0^*$ is only used in evaluating
$\textsf{MAC}_{k_0^*}()$. To prove the lemma, it suffices to show
that the simulator never evaluates and outputs  \textsf{MAC}$_{k_0^*}()$ with
input $C_i|S|y^*|\zeta^*|2$. Otherwise, it must be done  by an
instance $\Pi_i^{\ell _i}$ in $C_i$ in generating $Flow_3$ (recall
inputs for $\tau_0,\tau_1, \tau_2$  have different formats).  Hence,
since $C_i|S|y^*$ implies $\Pi_i^{\ell _i}$ samples
$x=\textsf{T}^*(\pi_i, y^*).$ It follows that $\ell _i=\ell _i^*$,
contradicting that $\Pi_i^{\ell ^*_i}$ is not partnered with
$\Pi_S^{\ell _S^*}.$ Hence, if
 $\Pi_i^{\ell _i}$ does not exist,
then  $\Pi_S^{\ell _S^*}$'s accepting $\tau_2^*$ implies a
\textsf{MAC} forgery,
 contradicting $\textsf{MAC}$ security!
$\hfill\square$

\begin{lemma} Recall {\bf Succ} be the success of ${\cal A}$ in the test session. Then,
$\Pr[{\bf Succ}\mid \neg\textsf{\em Non-Auth}]=1/2$ in $\Gamma_2$.
\label{le: succ}
\end{lemma}
{\em Proof. } Let $\Pi_U^{\ell _U^*}$ be the test instance and
$\textsf{pid}_U^{\ell ^*_U}=V$.  Let
$\textsf{sid}_U^{\ell _U^*}=C_J|S|y^*|\zeta^*$. Then, $\{U, V\}=\{J,
S\}.$ If $U=J$, then $V=S$ and (by Lemma \ref{le: part1}) there is
the unique partnered $\Pi_S^{\ell _S^*}$ for $\Pi_J^{\ell _J^*}$. If
$U=S,$ then $V=J$. In this case, if it does not exist a partnered
$\Pi_J^{\ell _J^*}$ in $C_J$ for $\Pi_S^{\ell _S^*}$, then
$\Pi_S^{\ell _S^*}$'s accepting $\tau_2^*$ implies
$\textsf{Non-Auth}_J$ event. Hence, under $\neg\textsf{Non-Auth}$
event, there is a partnered $\Pi_J^{\ell _J^*}$ for
$\Pi_S^{\ell _S^*}$ and by \textsf{Normal} condition it is unique. So
in any case, conditional on $\neg\textsf{Non-Auth}$, there is a
uniquely partnered $\Pi_V^{\ell _V^*}$ for $\Pi_U^{\ell _U^*}.$  Let
$(k_0^*, k_1^*)$ be the uniformly random keys defined to replace
$H_\theta(J, x^*)$ where $x^*=\textsf{T}^*(\pi_J, y^*).$ Let $b\in
\{0, 1\}, \alpha_1\in \{0, 1\}^\kappa$ be the random number in
\textsf{Test} oracle. We notice that in $\Gamma_2$,
$sk_U^{\ell _U^*}=k_1^*$ is taken uniformly random from $\{0,
1\}^{\kappa}.$ Let $\alpha_0=sk_U^{\ell _U^*}$. Let the randomness in
the whole game for $\Gamma_2$, except $k_1^*, b, \alpha_1$, be
denoted by $r$. Use $\textsf{View}_t({\cal A})$ to denote the
adversary view after the $t$th query. Then to prove the lemma, it
suffices to show that \textsf{View}$_t({\cal A})$ for each $t$ is
deterministic in $r, \alpha_b.$ We actually also show that
$\{\textsf{stat}_i^{\ell _i}\}_{(i, \ell _i)\ne (J, \ell _J^*), (S,
\ell _S^*)}$
 is also deterministic in $r, \alpha_b$.
Initially, \textsf{View}$_0({\cal A})$ is public parameters and
the conclusion trivially holds. Assume it is true for $t-1$ queries. Consider query $t.$

\vspace{0.05in} \noindent \textsf{Send}$(0, i, \ell _i, null).$ The randomness in sampling $x$ and the
randomness for $k_0$
is from $r.$ Hence, $C_i|y|\tau_0$ is deterministic in \textsf{View}$_{t-1}({\cal A})$
 and the randomness
$r$. \textsf{stat}$_i^{\ell _i}=C_i|S|y|k_0|k_1$.
When $(i, \ell _i)\ne (J, \ell _J^*)$, $k_1$ is determined by $r$. Hence, the conclusion holds after this query.

\vspace{0.05in} \noindent \textsf{Send}$(1, S, \ell _S,
C_i|y|\tau_0).$ Oracle first checks if $(i, y, *, *)\in
{\cal Q}$. If yes, extract $k_0$ from it and proceed normally (using
randomness in $r$ if needed).
 If no, compute $(k_0', k_1')=H_\theta(i, x)$ for $x=\textsf{T}^*(\pi_i, y)$ and proceed normally.
Notice the component $(i, y, k_0)$ in a record from ${\cal Q}$ is
computed using the randomness $r$; $\zeta$ is generated using $r$
too. $\theta$ is based on the randomness in the initialization of
$\Gamma_2$ and hence based on $r$ too. So adversary view in this
query is deterministic in \textsf{View}$_{t-1}({\cal A})$ and $r.$
If it outputs $Flow_2$, then $\textsf{stat}_S^{\ell _S}$ is updated
as $C_i|S|y|k_0|k_1$. By the uniqueness of $\ell _S^*$, when $(S,
\ell _S)\ne (S, \ell _S^*)$, $k_1$ is computing with $r$. Hence, the
conclusion holds after this query.

\vspace{0.05in}\noindent\textsf{Send}$(2, \cdot)$ and
\textsf{Send}$(3, \cdot)$ is deterministic in the view of ${\cal A}$
before the query and its session state. By the induction, the
conclusion holds after this query.

\vspace{0.05in}\noindent {\bf Reveal}$(i, \ell _i)$. This
query is $sk_i^{\ell _i}$. By the restriction on \textsf{Test} definition,
$\Pi_i^{\ell _i}\ne \Pi_S^{\ell _S^*}, \Pi_J^{\ell _J^*}$ and hence by induction,
its internal state is deterministic in
\textsf{View}$_{t-1}({\cal A})$  and $r, \alpha_b$. Since $sk_i^{\ell _i}$ is in
 his internal state, the conclusion holds after this query.

\vspace{0.05in} \noindent{\bf Corrupt}$(i)$. Upon this query $\pi_i$ as well
as $\{\textsf{stat}_i^{\ell _i}\}_{\ell _i}$
will be available to ${\cal A}$. Since $i\ne J, S$ by \textsf{Test} restriction,
 by induction, the conclusion holds after this query.

\vspace{0.05in} \noindent{\bf Test}$(u, \ell _u^*)$. Reply in this query is $\alpha_b$.
The conclusion holds trivially after this query.

As a summary, after any query, our conclusion holds. Hence, adversary view is
independent of $b.$  $\hfill\square$

\begin{lemma}
 $\Pr[\mbox{\bf Non-Auth}_i({\cal
A}, \Gamma_2)]\le \frac{Q_i}{|{\cal D}|}+negl(\kappa).$ \label{le:
non-auth}
\end{lemma}

\noindent {\em Proof. } To prove the lemma, we show how to simulate
$\Gamma_2$ when  the  randomness for $\{\pi_i\}_i$ is unfixed while the remaining randomness in the game is fixed. Let ${\cal D}_i$ be the probability
space for $\pi_i$ after each oracle query. We will simulate $\Gamma_2$ such that after each query,  the adversary view is identical for each
$(\pi_1, \cdots, \pi_n)\in {\cal D}_1\times {\cal
D}_2\times\cdots\times {\cal D}_n$.
Hence, given the adversary view, $(\pi_1, \cdots, \pi_n)$ is
uniformly distributed over ${\cal D}_1\times\cdots\times {\cal
D}_n.$

Initially, the adversary is given $\langle desc(\Lambda), \alpha(\theta)\rangle$ which is independent of $\pi_1, \cdots, \pi_n.$ Hence,
${\cal D}_1=\cdots={\cal D}_n={\cal D}.$
Assume this simulation is done for query $t-1.$ Consider query $t$, which is one of the following.

\vspace{0.05in} \noindent{\bf Send}$(0, i, \ell _i, null)$.
Oracle takes $y\leftarrow X$, $(k_0, k_1)\leftarrow \{0,
1\}^{2\kappa}$ and computes $\tau_0=\textsf{MAC}_{k_0}(C_i|S|y).$
Finally, update ${\cal Q}={\cal Q}\cup \{(i, y, k_0, k_1)\}$.
 The adversary view in this query is $C_i|y|\tau_0.$
  For any $\{\pi_j\}_{j=1}^n\in \prod_{j=1}^n {\cal D}_j$,
  the adversary view in the current query is identical.
 By induction assumption, after this query, if ${\cal D}_j, t=1, \cdots, n$ remains unchanged, the conclusion holds.
$ \textsf{stat}_i^{\ell _i}=C_i|S|y|k_0|k_1.$

 \vspace{0.05in} \noindent{\bf Send}$(1, S, \ell _S, C_i|y|\tau_0).$ \hspace{0.10in}
 Upon this, if $(i, y, k_0, k_1)\in {\cal Q}$,
 then (regardless of  the concrete value for $\pi_i$), the oracle will take $(k_0, k_1)$ from it and finish the remaining simulation
 in this query normally and all $\{{\cal D}_t\}$  remain unchanged. If $(i, y, k_0, k_1)\not\in {\cal Q}$, oracle will uses $\theta$ and $\pi_i$ to verify $\tau_0$  and
 announce the success of ${\cal A}$ if valid and reject otherwise. The analysis for this case is as follows.
 \begin{itemize}
\item[1.] {\bf $\tau_0$ is valid for the case  $\textsf{T}^*(\pi_i, y)\in L.$} \quad This case occurs only for at most
 one $\pi_i$ (denoted by $\pi_i(y)$) by Regularity Property {\bf R-2} of $(\textsf{T}, \textsf{T}^*).$
\item[2.] {\bf $\tau_0$ is valid for the case $\textsf{T}^*(\pi_i, y)\not\in L.$}
 \quad This event is a \textsf{Bad} event
in $\Gamma_2$ (negligible, see Lemma \ref{le: bad}). Since we already assume this event never occurs after Lemma
\ref{le: bad}, this case does not exist.
\end{itemize}
As a summery,  item 1 occurs (hence $\pi_i=\pi_i(y)$) with
probability at most $1/|{\cal D}_i|$ by induction assumption (since,
given $\textsf{View}_{t-1}({\cal A})$, vector $\{\pi_j\}_j$ is
uniform in $\prod_j{\cal D}_j$ and especially $\pi_i$ is uniform in
${\cal D}_i$); when item 1 does not occur, then   the
adversary view in this query is identical (i.e., \textsf{reject}) for any password setup:  take $\pi_i\in{\cal
D}_i\backslash\{\pi_i(y)\}$ and take $\pi_j\in {\cal D}_j$ for all $j\ne i$. Hence,
in this case,
 ${\cal D}_j$ for $j\ne i$ remain unchanged,
${\cal D}_i={\cal D}_i\backslash \{\pi_i(y)\}$ and
$\textsf{stat}_S^{\ell _S}=C_i|S|y|k_0|k_1$ is well defined.

\vspace{0.05in}\noindent  $\textsf{Reveal}, \textsf{Test}, \textsf{Send}(2, \cdots), \textsf{Send}(3, \cdots)$ are all processed only with a session state from
$\textsf{Send}(0, \cdot)$ oracle or $\textsf{Send}(1, \cdot)$ oracle, which is well defined as seen above. Hence, the simulation is perfect.

\vspace{0.10in} \textsf{Corrupt}$(i)$. \hspace{0.05in} In this case, $\pi_i$ is revealed and hence ${\cal D}_i$ is updated
 to a set of a single value. Notice that  $\{\textsf{stat}_i^{\ell _i}\}_{\ell _i}$ are consistent with all $\{\pi_j\}_j\in\prod_j{\cal D}_j$ by induction. Thus,  if
 we keep ${\cal D}_j$  unchanged for $j\ne i$, then
 the conclusion still holds.

Now we consider $\textsf{Non-Auth}_i$ event. It occurs at either
some $\Pi_i^{\ell _i}$ or $\Pi_S^{\ell _S}$ with
$\textsf{pid}_S^{\ell _S}=C_i.$ By Lemma \ref{le: part1}, it is
impossible to the former. For the latter, by Lemma \ref{le: part2},
it must hold that $(i, y, *, *)\not\in{\cal Q}$ in ${\bf Send}(1,
S, \ell _S, C_i|y|\tau_0)$ query and hence item 1 (i.e.,
$\pi_i=\pi_i(y)$) must occur (since item 2 is negligible and
ignored). It remains to calculate the probability $\pi_i=\pi_i(y)$
throughout the game. As analyzed above, it has a probability
$1/|{\cal D}_i|$, conditional on that previous queries with
$Flow_1=C_i|*$ do not have such an event. Hence, as a summery,
$\pi_i=\pi_i(y)$ occurs in the $\ell $th such a $\textbf{Send}(1, S,
\cdot, C_i|\cdot|\cdot)$ query with probability $\frac{|{\cal
D}|-1}{|{\cal D}|}\cdot\frac{|{\cal D}|-2}{|{\cal
D}|-1}\cdots\frac{1}{|{\cal D}|-{\ell -1}}=\frac{1}{|{\cal D}|}.$ We
claim that  there are at most $Q_i$ \textsf{Send}$(1, S, \cdot,
C_i|y|\cdot)$ queries for fixed $C_i$ such that $(i, y, *,
*)\not\in{\cal Q}$ with $\textsf{Client}(\Pi_S^{\ell _S})=C_i$. Indeed, although at
the beginning of theorem proof, we decompose {\bf Execute} into 4
{\bf Send}$(d,\cdot)$ queries, this treatment does not invalidate
the above statement: \quad in the special $\textbf{Send}(1, S,
\ell _S, C_i|y|\tau_0)$ query (decomposed from query $\textbf{Execute}(i,
\ell _i, S, \ell _S)$), $(i, y,
*, *)\in{\cal Q}$ was recorded by $\Pi_i^{\ell _i}$ in $\textbf{Send}(0, i, \ell _i,
null)$ (decomposed from the same \textsf{Execute} query). So {\bf Non-Auth}$_i$ does not occur to such a special \textsf{Send} query. Thus,
$\Pr[\mbox{\bf Non-Auth}_i({\cal A}, \Gamma_2)]\le\frac{Q_i}{|{\cal
D}|}.$ $\hfill\square$

\vspace{0.15in} We come back to the proof of theorem. Note that {\bf
Non-Auth}$_i$ and ${\bf Succ}$ both are in the view of adversary.
Hence, each of them are negligibly close between games $\Gamma_0,
\Gamma_1, \Gamma_2.$ By Lemmas \ref{le: succ} and \ref{le:
non-auth}, we conclude the theorem proof.  $\hfill\blacksquare$

\section{Persistency}
In this section, we show that our protocol is persistent against
the leakage of server key $\theta$. In our analysis,
 we model \textsf{MAC} as a random
oracle, which is reasonable (say, if we use HMAC).  We first introduce the following notion.
\begin{definition} $H_\theta:  \{0, 1\}^*\times X\rightarrow
\{0, 1\}^{2\kappa}$ is a tag-PHF and $F:  {\cal D}\times
X\rightarrow X$ is a deterministic function. $H_\theta$ is {\bf
locally 1-unique} w.r.t. $F$ if for any PPT adversary
${\cal A}$, the probability that the first $k$ bits of $H_\theta(z, F(\pi_1,
y))$ and $H_\theta(z, F(\pi_2, y))$ equal, is negligible,where $\pi_1, \pi_2$ are distinct and respectively goes over ${\cal D}$ and  $(z, y)\leftarrow
{\cal A}(\theta, \Theta, \pi_1, \pi_2)$.

\end{definition}

The persistency requires that  if the attacker runs
$T<\alpha\ell |{\cal D}|$ basic  steps for $\alpha<1/2$,  then the probability
for him to break the authentication w.r.t. $\ell$ clients, is small.  An authentication break occurs w.r.t. $C_i$ means
that either $\tau_1$ is accepted at {\bf Send}$(2, i, \ell_i,
Flow_2)$ while server $S$ never computes it, or $\tau_2$ is accepted
at {\bf Send}$(3, S, \ell_S, \tau_2)$ while $C_i$ never computes it. This intuitively  requires the knowledge of  $\pi_i$ to compute $k_0$ (hence $\tau_1, \tau_2, \tau_3$).
In our proof, we maintain and update  the candidate space ${\cal D}_i$ for  $\pi_i$ after each oracle query. We show that each query will either identify $\pi_i$ with probability $1/|{\cal D}_i|$ for a particular $i$, or remove one candidate of  $\pi_i$ from ${\cal D}_i$. But in any case, other ${\cal D}_j$ is not affected. Thus, an oracle query is similar to red ball experiment in Section 4:  it either hits a password $\pi_i$ (red ball) or remove one incorrect candidate (white ball) for $\pi_i$. From Theorem \ref{thm: prob}, we know that if there are at most $T<\alpha\ell |{\cal D}|$ coupons, then the probability to draw $\ell$ red balls in total is exponentially small.
We now proceed to a formal analysis.

\begin{theorem} Let $\textsf{\em MAC}: \{0, 1\}^\kappa\times \{0,
1\}^*\rightarrow \{0, 1\}^\kappa$ be a random oracle and
$H_\theta()$ is locally 1-unique with respect to $\textsf{\em T}^*.$
 Then, $\textsf{\em HPS-PAKE}$ is persistent, where assume one \textsf{\em MAC} evaluation is a basic step. \label{thm:
per}
\end{theorem}

\noindent{\bf Proof. } We first modify {\bf Send}$(0, \cdot)$ oracle
such that $x\leftarrow D(X\backslash L)$ (instead of $D(L)$). Since
$H_\theta(z, x)$ can be computed using $\theta$ (known), the revised
game can be simulated without difficulty. Thus, the probability to break authentication in these two
games differs negligibly. Hence, we only need to focus on the revised game. Assuming the
randomness $r$ for the whole game except $\{\pi_i\}$, is  fixed, we
show how to simulate {\bf Send} oracle and MAC oracle without
specifying $\{\pi_i\}.$ We denote ${\cal D}_i$ to be the candidate space for
$\pi_i$, given the current adversary view. We use
\textsf{view}$_t({\cal A})$ to denote the adversary view after $t$
    queries. Initially, ${\cal A}$ receives $\Theta,
desc(\Psi), \theta$, which is independent of $\{\pi_i\}.$ Hence,
given \textsf{view}$_0({\cal A})$, ${\cal D}_1=\cdots={\cal
D}_n={\cal D}.$ Assume the first $t-1$ queries are answered.
Consider query $t$ which is one of the following.

\vspace{0.05in} \noindent {\bf MAC} oracle. It maintains a
MAC list ${\cal L}$ consisting of records
 $(x, \textsf{MAC}(x)).$

\noindent{\bf Query} $m$ by Simulator. This query always has a
format $({\tt udef-}k_0, C_i|S|y|\rho)$ where $\rho=\epsilon$ (empty) or $\zeta|1$ or $\zeta|2$, where ${\tt udef-}k_0$ is the
random variable $k_0$ (dependent on the  random variable $\pi_i$), where
recall that the randomness other than $\{\pi_j\}$ is fixed.
Note that ${\tt udef-}k_0$
is determined if ${\cal D}_i$ has  a single element. By default, we assume that when $|{\cal D}_i|=1$, the simulator always realizes
${\tt udef-}k_0$ with the determined value.
 Upon the MAC query, it checks if it has been queried
before. If no, take $z\leftarrow \{0, 1\}^\kappa$ and add $(({\tt udef-}k_0, C_i|S|y|0), z)$ into ${\cal L}.$ In
any case, return $z$ for $({\tt udef-}k_0, C_i|S|y|\rho), z)\in
{\cal L}$ as the reply. Note that this query does not change $\{{\cal D}_j\}$. That is, the simulation is consistent for any assignment $\{\pi_j\}\in \prod_j{\cal D}_j.$

\vspace{0.05in} \noindent{\bf Query} $m$ by ${\cal A}$. \quad Upon
query $m$, if $m$ was queried before, find $y$ such that $(m, y)\in
{\cal L}$ and return $y$.  If $m$ was not queried before, do the
following. If $m$ can be parsed into a format $(u, s)$ where
$s=C_i|S|y|\rho$ for $\rho=\epsilon$ (empty) or $\zeta|1$ or $\zeta|2$, then check whether  there exists some $\pi(i, y)\in {\cal D}_i$
(unique if any, by assumption on $H_\theta()$) s.t. $(u,
*)=H_\theta(i, \textsf{T}^*(\pi(i, y), y))$. If $\pi(i, y)$ exists, check whether
$\pi_i=\pi(i, y)$ (in this case, `=' occurs with probability
$1/|{\cal D}_i|$ since any $\{\pi_j\}_j\in\prod_j{\cal D}_j$ gives
the same adversary view).  If yes, ${\cal D}_i=\{\pi(i, y)\}$ and set  $\textsf{udef-}k_0$ in record $(\textsf{udef-}k_0, C_i|S|y'|\rho)$ of  ${\cal L}$ by
the first $k$bits of $H_\theta(i, \textsf{T}^*(\pi(i, y), y'))$;
 otherwise, set ${\cal D}_i={\cal D}_i\backslash \{\pi(i, y)\}.$
When query $m$ was not recorded in ${\cal L}$, take $z\leftarrow \{0,
1\}^\kappa$ (using random tape $r$) and add $(m, z)$ into a list
${\cal L}$. In any case, return $z$ for $(m, z)\in {\cal L}.$

Our \textsf{MAC} simulation above has the property that if any
$\{\pi_i\}\in \prod_i{\cal D}_i$ (also realizing  $\textsf{udef}-k_0$ based on this assignment accordingly) before the MAC query is consistent with adversary view, then after the MAC query, this still holds for updated $\{{\cal D}_i\}.$

\vspace{0.05in}\noindent{\bf Send}$(0, i, \ell _i, null).$ Upon this,
take $y\leftarrow X$. Assume no query $(*, C_i|S|y)$ was previously issued to
\textsf{MAC}, which is violated with probability $|{\cal L}|/|X|$
(tiny and ignored!). Query $({\tt udef}-k_0, C_i|S|y)$ to \textsf{MAC} oracle and
when receiving the reply $z$, define $\tau_0=z.$
Finally, send $C_i|y|\tau_0$ to ${\cal
A}$.

\vspace{0.05in}\noindent{\bf Send}$(1, S, \ell _S, C_i|y|\tau_0).$
 Upon this, query $({\tt udef}-k_0, C_i|S|y)$ to \textsf{MAC} oracle and when receiving the reply $z$, $\tau_0$ is accepted
if and only if $\tau_0=z.$ If $\tau_0=z$, normally generate $Flow_1$ by querying
$({\tt udef-}k_0, C_i|S|y|\zeta)$ to \textsf{MAC} oracle for computing $\tau_1$.
 Finally send out $S|\tau_1|\zeta$.

\vspace{0.05in} \noindent {\bf Send}$(2, i, \ell _i,
S|y|\tau_1|\zeta).$ \quad Upon this, verify $\tau_1$ by querying
$({\tt udef-}k_0, C_i|S|y|\zeta|1)$ to $\textsf{MAC}$ oracle and if accepted, generate and send out $\tau_2$ by
querying $({\tt udef-}k_0, C_i|S|y|\zeta|2)$ to \textsf{MAC} oracle.

\vspace{0.05in} \noindent {\bf Send}$(3, S, \ell _S, \tau_2).$
Verify $\tau_2$ by a query  $({\tt udef-}k_0, C_i|S|y|\zeta|2)$ to \textsf{MAC} oracle.

\vspace{0.05in} By the definition of \textsf{MAC}, after each query,
the adversary view will be consistent with any $\{\pi_j\}_j\in
\prod_j{\cal D}_j$. Our simulation is perfect consistent with the
real game.

It important to know that each \textsf{Send} oracle only does not change $\prod_j{\cal D}_j$: \quad it only involves a MAC query from Simulator which does not change $\prod_j{\cal D}_j$ and the remaining code in \textsf{Send} oracle does not change it either. Now violation of authentication w.r.t. a client $C_i$ occurs only in
two cases:

\vspace{0.05in}\noindent \textbullet\quad  In {\bf Send}$(2, i, \ell _i,
S|y|\tau_1|\zeta)$, where $\tau_1$ is accepted while tuple $({\tt
udef}-k_0, C_i|S|y|\zeta|1)$ was not queried to \textsf{MAC} oracle
by {\bf Simulator} before this \textsf{Send} query. By treatment of
\textsf{MAC} oracle, when $|{\cal D}_i|\ge 2,$  no $(\pi, C_i|S|y|\zeta|1)$ for any $\pi\in {\cal D}_i$ is queried to $\textsf{MAC}$; otherwise,
either $|{\cal D}_i|=1$ (for case $\pi_i=\pi$) or $\pi$ was removed from ${\cal D}_i$ (for case $\pi_i\ne \pi$).  Hence, given adversary view, $\textsf{MAC}({\tt
udef}-k_0, C_i|S|y|\zeta|1)$ is random in $\{0, 1\}^\kappa$ and hence
$\tau_1$ will be rejected (ignore the  probability $2^{-\kappa}$ of acceptance), when  $\pi_i$ is set to any value in ${\cal D}_i$. This also implies that after this query,
 $\{{\cal D}_j\}$ remains unchanged since for any assignment $\{\pi_j\}\in \prod_j{\cal D}_j$  the adversary view in this query is identical: reject.  When $|{\cal D}_i|=1$, $\tau_1$ is of course  accepted with probability  at most $1/|{\cal D}_i|=1.$

\vspace{0.05in}\noindent \textbullet\quad In  {\bf Send}$(3, S, \ell_S, \tau_2)$ with $\textsf{pid}_S^{\ell_S}=C_i$, {\bf Simulator} has never  queried
$({\tt udef-}k_0, C_i|S|y|\zeta|2)$ to $\textsf{MAC}$ oracle but $\tau_2$ is valid. The analysis is similar to \textsf{Send}$(2, \cdot)$ above.

\vspace{0.05in} Now we evaluate $\Pr[{\bf Succ}({\cal A})]$.  From the above analysis, authentication breaks occurring w.r.t.  $\ell$ clients implies that $|{\cal D}_i|=1$ for these clients.  On the other hand, we have shown that ${\cal D}_i$ is reduced only when ${\cal A}$ makes some special \textsf{MAC} queries $(u, s)$ that defines
$\pi(i, y)$: if $\pi(i, y)=\pi_i$ with probability $1/|{\cal D}_i|$; otherwise, ${\cal D}_i={\cal D}_i\backslash \{\pi(i, y)\}$. Now we can build  red ball experiment out of this event:
$\pi_i$ is red ball and $\pi(i, y)$ is a pick at Box i. $\pi(i, y)$ hits the red ball with probability  $1/|{\cal D}_i|$; otherwise, Box i eliminates one white ball $\pi(i, y).$
Defining $\pi(i, y)$ involves at least one MAC computation. Hence, one pick costs at least one basic step.  By Theorem \ref{thm: prob}, within $T<\alpha\ell |{\cal D}|$ picks, $\ell$ red balls are selected with probability  at most
 by $\exp(-2\ell(0.5-\alpha)^2).$
$\hfill\blacksquare$

\section{Realization by  Revised Cramer-Shoup Hash Proof System} \label{sec: concrete}

In this section, we realize \textsf{HPS-PAKE} framework using a tag-HPS, revised Cramer-Shoup  hash proof system
\cite{KD04,CS01}.

\vspace{0.05in}\noindent\textbullet\quad \textbf{Hard Subset
Membership Problem}.\hspace{0.10in}  Sample a prime
$p=2q+1$ where $q$ is also a large prime. Let $\mathbb{G}$ be the
prime group of $\mathbb{Z}_p^*$ of order $q.$ Take $g_1,
g_2\leftarrow \mathbb{G}$. The set $X=\{(g_1^{r_1}, g_2^{r_2})\mid
r_1, r_2\in\mathbb{Z}_q\}.$ Language $L$ is defined as $L=\{(g_1^r,
g_2^r)\mid r\in \mathbb{Z}_q\}.$ The witness for $(g_1^r, g_2^r)\in
L$ is $r$. $D(L)$ and $D(X\backslash L)$ are uniform distributions
over $L$ and $X\backslash L$, respectively. Witness set
$W=\mathbb{Z}_q$.  {\bf NP}-relation $R$ is  defined as $R=\{(r,
(u_1, u_2))\mid u_1=g_1^r, u_2=g_2^r, r\in W\}$. Hence, the
description $desc(I_\kappa)=(g_1, g_2, p).$ This is a hard
subset membership problem by Decisional Diffie-Hellman (DDH)
assumption in $\mathbb{G}$. \remove{Here DDH assumption states: \quad
 no polynomial algorithm can distinguish  distributions \textsf{DH} and \textsf{Rand}, where
 $\textsf{DH}=(g_1, g_2,
g_1^x, g_2^{x})$ for $x\leftarrow  \mathbb{Z}_{q}$ and
$\textsf{Rand}=(g_1, g_2, g_1^x, g_2^{y})$ for $x, y\leftarrow
\mathbb{Z}_{q}$.}

\vspace{0.05in}\noindent\textbullet\quad {\bf Tag-based
Projective Hash Function $\Psi$.} \hspace{0.10in} Let $S=\mathbb{G}^2$
and $G=\{0, 1\}^{2\kappa}$.  Let key space ${\cal K}=\{(a_1, a_2,
b_1, b_2)\mid a_1, a_2, b_1, b_2\in \mathbb{Z}_q\}.$
$\Theta=\alpha(\theta)=(\Theta_1, \Theta_2)=(g_1^{a_1}g_2^{a_2},g_1^{b_1}g_2^{b_2})$, for
$\theta=(a_1, a_2, b_1, b_2)\in {\cal K}.$  Let $h_\lambda$ be a
collision resistent hash function from $\{0, 1\}^*$ to $\mathbb{Z}_q$, indexed by $\lambda\leftarrow
\{0, 1\}^\kappa$. Let \textsf{KDF} is a key
derivation function (e.g., the least half bits of the input) and is
not used   in the original HPS \cite{KD04,CS01}.  For $(u_1, u_2)\in X$ and a tag $z$,
define $H_\theta(z, (u_1,
u_2))=\textsf{KDF}(u_1^{a_1+b_1\tau}u_2^{a_2+b_2\tau}),$ where
$\tau=h_\lambda(z, u_1, u_2).$ If $(u_1, u_2)=(g^r_1, g_2^r)$, then
$H_\theta(z, u_1, u_2)=
\textsf{KDF}(u_1^{a_1+b_1\tau}u_2^{a_2+b_2\tau})$

$=\textsf{KDF}((\Theta_1\Theta_2^\tau)^r).$ So $\Psi$ is
a projective hash function and $desc(\Psi)=(g_1, g_2, \lambda, p).$ By Lemma \ref{le: cu2} below,
 $\Psi$ is also computational universal$_2$.

\vspace{0.05in}\noindent\textbullet\quad {\bf Regular Transformation
Pair} ($\textsf{T}, \textsf{T}^*$): \hspace{0.05in} For $\pi\in
{\cal D}$ and $(u_1, u_2)\in X,$ define $\textsf{T}(\pi, (u_1,
u_2))=(u_1, u_2 g_2^\pi)$ and $\textsf{T}^*(\pi, (u_1, u_2))=(u_1,
u_2g_2^{-\pi}).$ Evidently, regularity property {\bf
R-1} is  satisfied. In addition, property {\bf R-2} is satisfied as
long as no $\pi_1, \pi_2\in {\cal D}$ s.t. $\pi_1\equiv\pi_2\
(\mbox{mod} p)$, which is evident when ${\cal D}=\{1, \cdots, N\}$
for $N<q.$

\begin{lemma} If $h_\lambda$ is collision-resistant, then $\Psi$ must be  computational universal$_2.$ \label{le: cu2}
\end{lemma}
The proof is similar to \cite[Lemma 6.3]{HK07} and omitted here.

\vspace{0.15in}\noindent {\bf Security. }  Let \textsf{HPS$_{cs}$-PAKE}
denote \textsf{HPS-PAKE} realized by the above tag-HPS. From Theorem \ref{thm: secure}, it is secure.
\remove{
\begin{corollary} $h_\lambda$ is collision-resistant. Then, under {\em decisional Diffie-Hellman} assumption,
\textsf{\em HPS$_{cs}$-PAKE} is secure. \label{co: cs}
\end{corollary}}

\vspace{0.05in} \noindent {\bf Persistency. } Now we consider the
persistency of \textsf{HPS$_{cs}$-PAKE}. By Theorem
\ref{thm: per}, we only need to show that $H_\theta(z, x)$ is {\em
locally 1-unique}, which is seen in the following lemma.

\begin{lemma}  If $h_\lambda$ is a random oracle,  $\textsf{\em dist}[\textsf{\em KDF}(V), U_\kappa]=\Delta$ so that $(\Delta+2^{-\kappa})N^2=negl(\kappa)$, where $V, U$ are uniform over $\mathbb{G}, \{0, 1\}^\kappa$ respectively.
 Then,  $H_\theta()$ is locally 1-unique with respect to $\textsf{\em T}^*$.
\end{lemma}

\noindent{\bf Proof. } Since $b_2$ is
uniform over $\mathbb{Z}_q$, we ignore the probability
$b_2=0.$ Let $(z^*, x_1^*, x_2^*)$ be the output of ${\cal
A}.$  For any distinct  $\omega_1, \omega_2\in [N],$ let $A=H_\theta(z^*, \textsf{T}^*(\omega_1, x_1^*, x_2^*))
={x^*}_1^{a_1}{x^*}_2^{a_2}g_2^{-\omega_1 a_2}\cdot
({x^*}_1^{b_1}{x^*}_2^{b_2}g_2^{-b_2\omega_1})^{\tau_1}$,
$B={x^*}_1^{a_1}{x^*}_2^{a_2}g_2^{-\omega_2 a_2}\cdot
({x^*}_1^{b_1}{x^*}_2^{b_2}g_2^{-b_2\omega_2})^{\tau_2},$
 where $\tau_1=h_\lambda(z^*, x^*_1, x^*_2g_2^{-\omega_1})$ and $\tau_2=h_\lambda(z^*, x^*_1, x^*_2g_2^{-\omega_2}).$
 As $q>N$, $\tau_1$ and $\tau_2$ are independent (in $\mathbb{Z}_q$) and
$$({x^*}_1^{b_1}{x^*}_2^{b_2}g_2^{-b_2\omega_1})/({x^*}_1^{b_1}{x^*}_2^{b_2}g_2^{-b_2\omega_2})=g_2^{b_2(\omega_2-\omega_1)}$$ has an order of $q$. Thus,
either $B$ or $A$ is uniformly distributed over $\mathbb{G}$. Assume $B$ has an order of $q$.  From independence between $\tau_1$ and $\tau_2$, $B$ is uniformly random over $\mathbb{G}$ for fixed $A$. So by calculation  the first $\kappa$ bits of $\textsf{KDF}(B)$ and $\textsf{KDF}(A)$ equal with probability $\le 2\Delta+2^{-\kappa}$. Since there are $N(N-1)/2$ pairs of $(\omega_1, \omega_2)$, by assumption, the lemma follows.  $\hfill\blacksquare$

\vspace{0.10in} \noindent{\bf Efficiency. }
\hspace{0.05in} Client's cost is dominated by 4
exponentiations for  $y=(g^r_1, g_2^{r+\pi})$ and $(k_0, k_1)=\textsf{KDF}((\Theta_1\Theta_2^\tau)^r).$ Server's cost is dominated by
2 exponentiations for $(k_0, k_1)=\textsf{KDF}(u_1^{a_1+b_1\tau}u_2^{a_2+b_2\tau})$ where $y=(u_1, u_2g_2^{\pi_i})$ (note he can store $g_2^{\pi_i}$).
Here we did not count the verification of $y\in \mathbb{G}$ by $S$ which needs one more exponentiation. However, we can use a recent technique (from our separate paper) to slightly modify the protocol so that  we can avoid  the verification by exponentiation. The modification for \textsf{HPS$_{cs}$-PAKE} is as follows.
 In $Flow_1$, instead of sending $y=(g_1^r,
g_2^{r+\pi_i})$, Client $i$ computes  $y':=(y_1', y_2'):=(g_1^{r/2},
g_2^{(r+\pi)/2})$ and let $y=({y_1'}^2, {y'}_2^2)$ and replace $y$ in
the original $Flow_1$ message by $y'.$ The remaining specification for Client is
unchanged. Correspondingly, Server computation
is as follows. It first recovers $y=({y'}_1^2, {y'}_2^2)$ from $y'$
when receiving $Flow_1$ and the remaining specification in Server is
unchanged. Denote the modified protocol by
\textsf{HPS$_{cs}^*$-PAKE}. The cost for client and server each
increases by 2 squarings, which is tiny. Then, the security of
\textsf{HPS$_{cs}$-PAKE} implies the security of
\textsf{HPS$_{cs}^*$-PAKE}. The proof uses  the fact that for
$y\in\mathbb{G},$ it holds that  $\sqrt{y}=y^{(q+1)/2}.$ The
security of \textsf{HPS$_{cs}^*$-PAKE} is obtained by proving that
if there is an adversary ${\cal A}'$ against
\textsf{HPS$_{cs}^*$-PAKE} with success probability $prob$, then
there exists an adversary \textsf{HPS$_{cs}$-PAKE} with the same
success probability. The setup of these two protocols are the same.
So when ${\cal A}$ receives the setup parameter $(desc(\Phi),
\Theta)$, it forwards to ${\cal A}'$. Upon {\bf Send} query from
${\cal A}'$, the strategy of ${\cal A}$ is to forward the query from
${\cal A}'$ to his own challenger and relay the reply from the
latter back to ${\cal A}',$ except $y$ in $Flow_1$ of
\textsf{Send}$(1, \cdot)$ query is replaced by $y'=\sqrt{y}.$ For
remaining queries ${\bf Reveal}, {\bf Corrupt}(i), {\bf Test}$ from
${\cal A}'$, ${\cal A}$ forwards it to his own challenger and
replays the reply back to ${\cal A}'.$ From this strategy, we know that whatever ${\cal A}'$ breaches \textsf{HPS$_{cs}^*$-PAKE}, ${\cal A}$ can do the same to
$\textsf{HPS$_{cs}$-PAKE}.$ Hence, the security of \textsf{HPS$_{cs}^*$-PAKE} follows. Details are omitted here.

\vspace{0.15in}

\noindent{\bf Appendix A.  \hspace{0.10in} Proof of Lemma \ref{le: CU2}}

\vspace{0.05in} \noindent{\bf Proof. } Use $\Re_c$ to denote $\Re$ when the
challenge bit is $c.$  It suffices to show that $\Pr[{\cal
A}(\Re_0)=1]=\Pr[{\cal A}(\Re_1)=1]+negl(\kappa).$ Let
$\Re_0^{\ell }$ denote the variant of $\Re_0$, where the first $\ell $
${\tt Challenge}$ queries are answered as in $\Re_1$ while the
remaining such queries are answered as in $\Re_0.$ Let $\sharp$ of
${\tt Challenge}$ queries be bounded by $N.$ Then, $\Re_0^0=\Re_0$
and $\Re_0^N=\Re_1.$  If the lemma is violated by ${\cal A}$, then
by hybrid argument, there exists $\ell $ such that $|\Pr[{\cal
A}(\Re_0^{\ell -1})=1]-\Pr[{\cal A}(\Re_0^{\ell })=1]|$ is
non-negligible. Let $\hat{\Re}_0^i, i=\ell -1, \ell $ be the variant
of $\Re_0^i$ such that in the $\ell $th ${\tt Challenge}$ query,
$x\leftarrow X\backslash L$ (instead of  $x\leftarrow L$),
where correspondingly $H_k(z, x)$ is computed using $k$. By reduction
to the hardness of ${\cal I}$, we have $\Pr[{\cal A}(\Re_0^i)=1]
=\Pr[{\cal A}(\hat{\Re}_0^i)=1]+negl(\kappa).$ Hence, $\Pr[{\cal
A}(\hat{\Re}_0^{\ell-1} )=1] -\Pr[{\cal A}(\hat{\Re}_0^{\ell})=1]$ is
non-negligible. We build an adversary ${\cal D}$  that uses  ${\cal
A}$ to break computationally universal$_2$ of $\Psi.$ Upon
public key $pk=(\alpha(k), desc(\Psi))$, ${\cal D}$
invokes ${\cal A}$ with $pk$ and simulates $\hat{\Re}_0^\ell $ with
it as follows. He defines $c$ to be the hidden bit in his challenge key $K_c$ (parsed as $(a^*_{c}, s^*_{c})$ in this proof).

\begin{itemize}
\item[\textbullet] \textsf{$i$th {\tt Challenge} Query with $z$ from ${\cal A}$. } \hspace{0.10in}  If
$i\ne \ell $,  take $x\stackrel{w}{\leftarrow} D(L)$ and compute
$(a_{0}, s_{0})=H_k(z, x)$ using $w$. The remaining simulation in this
query is normal as in $\hat{\Re}_0^\ell$. If $i=\ell $, he takes $x^*\leftarrow D(X\backslash L)$ and sets $(z, x^*)$ to be
his test pair $(z_2, x_2)$. In turn, he will receive $K_c$ (parsed as $(a^*_{c}, s^*_{c})$) and then  he forwards to ${\cal A}$. Then, he
updates
$\Theta=\Theta\cup\{(z, x^*, a^*_{c}, s^*_c)\}$.

\item[\textbullet] \textsf{{\tt Compute} Query} $(z, x, \sigma, m).$
\hspace{0.10in} If $(z, x, a', s')\in \Theta$ for some $a', s'$, verify
$\sigma$ using $a'$;
 otherwise, he issues {\bf Evalu} query to his challenger with
$(z, x)$ and in turn receives $(a, s)$. If $(a, s)=\perp$ (hence $x\not\in L$) or $\sigma\ne \textsf{MAC}_a(m)$, he outputs $\perp$; otherwise,
he outputs $(a, s)$.
\end{itemize}
At the end of game, ${\cal D}$ outputs whatever ${\cal A}$ does.

Denote the simulated game of ${\cal D}$ with bit $c$ by
$\bar{\Re}_0^{\ell -c}.$ Then $\bar{\Re}_0^{\ell -c}$ is identical
to $\hat{\Re}_0^{\ell -c}$, except in the case of $x\not\in L$ in
${\tt Compute}$ query. In this case, the challenger of ${\cal D}$
returns $(a, s)=\perp$ and ${\cal D}$ will output $\perp$ too while
in $\hat{\Re}_0^{\ell -c}$, $\sigma$ will be verified using $a$ in
$(a, s)=H_k(x)$ and (if valid) $(a, s)$ is returned. Hence, inconsistency between the two games occurs only if
 the following event occurs to some \textsf{Compute} query $(z, x, \sigma, m)$ in $\bar{\Re}_0^{\ell -c}$:   $(z, x, *, *)\not\in\Theta$
and $x\not\in L$ but $\sigma=\textsf{MAC}_a(m)$.  Denote this
event by $\textsf{E}$.
we have that   $|\Pr[{\cal A}(\hat{\Re}_0^{\ell
-c})=1]-\Pr[{\cal A}(\bar{\Re}_0^{\ell -c})=1]|\le
\Pr[\textsf{E}(\bar{\Re}_0^{\ell -c})]$. We claim that
$\Pr[\textsf{E}(\bar{\Re}_0^{\ell -c})]=negl(\kappa), c=0, 1;$
otherwise, computational universal$_2$  of $\Psi$ can be broken by
adversary ${\cal D}'$ as follows. W.O.L.G, assume
$\Pr[\textsf{E}(\bar{\Re}_0^{\ell })]$ is non-negligible. Upon
receiving $pk$, ${\cal D}'$ simulates $\bar{\Re}_0^\ell $ by playing
the role of ${\cal D}$ and the challenger of ${\cal D}$, where $pk$
is the public key, except the evaluation of $H_k(z, x)$ is done
under his own challenger's help. Specifically, for the $i$th {\tt
Challenge} query for $i\ne \ell $, he can take $x\leftarrow L$ and
compute $H_k(z, x)$ with $w$ himself ; For the $\ell $th {\tt
Challenge} query, he takes $x^*\leftarrow X\backslash L$ and asks
his challenger to evaluate $H_k(z, x^*)$ as the first challenge (i.e, $(z_1, x_1)$ in Definition \ref{def: phf});
upon a {\tt Compute} query $(z, x, \sigma, m)$, he asks his own
challenger with $(z, x)$ and in turn he will receive $(a, s)=\perp$
if $x\not\in L$; $H_k(z, x)$ otherwise.  In case of the former, he
records $(z, x)$ in to a list ${\cal L}$ and reject normally (as in
$\bar{\Re}_0^{\ell-c}$); in case of the latter, answer the query
using the received $H_k(z, x)$ normally.
 The remaining simulation is normal. This simulation is perfectly consistent
 with $\bar{\Re}_0^{\ell-c}$ for both cases $c=0$ and $1$.
 At the end of game,
if $c=1$ (since we only consider $\bar{\Re}_0^{\ell}$, not
$\bar{\Re}_0^{\ell-1}$), he outputs 0/1 randomly; otherwise, he
takes $(z^*, y^*)$ randomly from ${\cal L}$ and ask $(z^*, y^*)$ as
his test challenge (i.e., $(z_2, x_2)$ in Definition \ref{def: phf}). In turn he will receive $(a_b^*, s_b^*)$,
where $(a_0^*, s_0^*)=H_{k}(z^*, y^*)$ or $(a_1, s_1)\leftarrow \{0,
1\}^{2\kappa}.$ Then he reviews all the {\tt Compute} queries in ${\cal L}$ with
forms $(z^*, y^*, \sigma, m)$ for any $\sigma, m$ and denote event
$\sigma=\textsf{MAC}_{a_b^*}(m)$ by $inc$. In case of $inc$, output
0; otherwise output 1. Note if $b=1$, then $inc$ occurs to $y^*$
negligibly by ungorgeability of $\textsf{MAC}$. If $b=0$, then $inc$
event is \textsf{E} event in $\bar{\Re}_0^\ell $ occurs to $(z^*,
y^*)$. Since any $\textsf{E}$ event must occur to some $(z, x)$ in
${\cal L}$, $inc$ occurs in ${\cal D}$'s algorithm for $b=0$ with
probability at least $\Pr[\textsf{E}(\bar{\Re}_0^\ell )]/|{\cal
L}|,$ non-negligible. The non-negligible gap of the two cases
implies non-negligible advantage of ${\cal D}'$, contradiction.
Hence, $\Pr[{\cal A}(\bar{\Re}_0^\ell )=1]-\Pr[{\cal
A}(\bar{\Re}_0^{\ell -1})=1]$ is non-negligible, which is the success advantage of ${\cal D}$, contradiction.
$\hfill\blacksquare$

\vspace{0.10in}

\noindent{\bf Appendix B.  \hspace{0.10in} Proof of Lemma \ref{le: reduce}}

\vspace{0.05in}\noindent{\bf Proof. }  Use $Left$ and $Right$ to denote the left
and right side of Eq. (\ref{eq: reduce}) respectively. First of all,
we show ${Left}\ge Right$ by presenting an algorithm ${\cal A}_0$
achieving $Right.$ ${\cal A}_0$ simply draws the ball from Box 1
until the red ball is picked. Then, he turns  to  Box 2 using the
same strategy, then Box 3, $\cdots$. If he draws a red ball from
Box $\ell$ before $t$ picks are used up, he succeeds; otherwise, he
fails. Let the red ball in Box $i$ be obtained by using  $x_i$ picks. Then, it is simple to verify that  $x_i\leftarrow [a_i].$ Hence,  the success probability of ${\cal
A}_0$ is exactly the right side of Eq. (\ref{eq: reduce}).

It remains to show that $Left\le Right.$ When $\ell=0$, the
conclusion holds trivially since both sides are 1. Assume $\ell\ge
1$. When $n=1$, two sides of Eq.  (\ref{eq: reduce}) equal
$\min\{t/a_1, 1\}$ for the (only) case $\ell=1$.  For $n\ge 2$ and
$\ell \ge 1.$  we use induction on $t.$ Note $\Theta_{t, n, \ell
}(a_1, \cdots, a_n)$ can always be achieved by a deterministic
algorithm by computing the maximum success probability over the randomness of ${\cal A}$. Hence, we assume a deterministic ${\cal A}$ achieves it.
When $t=0$, two sides of Eq. (\ref{eq: reduce}) are zero. The
conclusion holds trivially. When $t=1$, assume the first box chosen
by ${\cal A}$ is $j.$ Then
\[
\begin{array}{ll}
&\Theta_{1, n, \ell  }(a_1, \cdots, a_n)\\
=&{a_j^{-1}}\cdot \Theta_{0, n, \ell-1}(a_1, \cdots, a_{j-1}, 0, a_{j+1}, \cdots, a_n)\\
  & +(1-{a_j^{-1}})\Theta_{0, n, \ell}(a_1, \cdots, a_{j-1}, a_j-1, a_{j+1}, \cdots, a_n)\\
  =&{a_j^{-1}}\cdot \Theta_{0, n-1, \ell-1}(a_1, \cdots, a_{j-1}, a_{j+1},
 \cdots, a_n)\\
  & +(1-{a_j^{-1}})\Theta_{0, n, \ell}(a_1, \cdots, a_{j-1}, a_j-1, a_{j+1}, \cdots, a_n)
 \end{array}
\]

If $\ell=1$, then this gives $\Theta_{1, n, \ell  }(a_1, \cdots,
a_n)=a_j^{-1}\le a_1^{-1}=Right$. Hence, $Left\le Right$.

If $\ell\ge 2,$ since
 $\Theta_{0, n-1, \ell-1}(a_1, \cdots, a_{j-1}, a_{j+1},
 \cdots, a_n)=0$ and $$\Theta_{0,    n, \ell}(a_1, \cdots, a_{j-1}, a_j-1, a_{j+1}, \cdots, a_n)=0,$$ we have that $\Theta_{1, n, \ell  }(a_1, \cdots, a_n)=0$.
In addition, since $x_1+\cdots+x_\ell\ge \ell>1$, $Right=0$. Hence, $Left=Right$.

Now assume $Left\le Right$  for $t-1$, which implies
$Left=Right$ for $t-1$ since $Left\ge Right$ is proven at the
beginning.
 We
consider $t$ ($t\ge 2$). Assume the first box chosen by ${\cal A}$
is $j$. Then,
\[
\begin{array}{ll}
&\Theta_{t, n, \ell  }(a_1, \cdots, a_n)\\
=&{a_j^{-1}}\cdot \Theta_{t-1, n, \ell-1}(a_1, \cdots, a_{j-1}, 0, a_{j+1},
 \cdots, a_n)\\
& +(1-{a_j^{-1}})\Theta_{t-1, n, \ell}(a_1, \cdots, a_{j-1}, a_j-1, a_{j+1}, \cdots, a_n)\\
  =&{a_j^{-1}}\cdot \Theta_{t-1, n-1, \ell-1}(a_1, \cdots, a_{j-1}, a_{j+1},
 \cdots, a_n)\\
 & +(1-{a_j^{-1}})\Theta_{t-1, n, \ell}(a_1, \cdots, a_{j-1}, a_j-1, a_{j+1}, \cdots, a_n)
 \end{array}
\]
There are two cases.

\vspace{0.05in}\noindent {\bf Case  $a_j=1$: } \quad In this case, we have
$\Theta_{t, n, \ell  }(a_1, \cdots, a_n)$ =$\Theta_{t-1, n-1,
\ell-1}(a_1, \cdots, a_{j-1}, a_{j+1},
 \cdots, a_n).$ Let  $a_1^*, \cdots, a_{\ell-1}^*$ be $\ell-1$ smallest numbers among $\{a_1, \cdots, a_n\}\backslash \{a_j\}.$
By induction,
\begin{equation}
\Theta_{t-1, n-1, \ell-1}(a_1, \cdots, a_{j-1}, a_{j+1},
 \cdots, a_n)=\Pr\big{[}x_1^*+\cdots+x_{\ell-1}^*\le t-1:\quad x_i^*\leftarrow [a_i^*]\big{]}.\end{equation} 

\noindent If $j>\ell,$ then $a_1=\cdots=a_\ell=1$ as $a_1\le a_2\le\cdots\le a_n.$ Hence, $(a_1^*,\cdots, a_{\ell-1}^*)$ equals
$(a_1, \cdots, a_{\ell-1}).$ Therefore,

$\Pr\big{[}\sum_{i=1}^{\ell-1}x_i^*\le t-1:\quad x_i^*\leftarrow [a_i^*]\big{]}=
\Pr\big{[}\sum_{i=1}^{\ell-1}x_i\le t-1:\quad x_i\leftarrow [a_i]\big{]}.$ Since $a_\ell=1$, it follows that  $x_\ell=1$ always holds when  $x_\ell\leftarrow [a_\ell].$ So
$\Pr\big{[}\sum_{i=1}^{\ell-1}x_i\le t-1:\quad x_i\leftarrow [a_i]\big{]}=
\Pr\big{[}\sum_{i=1}^{\ell}x_i\le t:\quad x_i\leftarrow [a_i]\big{]}.$  The induction holds in this case.

\noindent If $j\le \ell,$ then $\{a_1^*, \cdots, a_{\ell-1}^*\}=\{a_1, \cdots, a_{j-1}, a_{j+1}, \cdots, a_\ell\}$. Hence,  
$$
\begin{array}{lll}
&&\Pr\big{[}\sum_{i=1}^{\ell-1}x_i^*\le t-1:\quad x_i^*\leftarrow [a_i^*]\big{]}\\
&=&
\Pr\big{[}\sum_{1\le i\le \ell, i\ne j}x_i\le t-1:\quad x_i\leftarrow [a_i]\big{]}\\
&=& \Pr\big{[}\sum_{i=1}^\ell x_i\le t:\quad x_i\leftarrow [a_i]\big{]},
\end{array}
$$
where the last `=' holds since $a_j=1$ and hence $x_j=1$ holds always. Hence, the induction holds in this case too.

\vspace{0.05in}\noindent{\bf Case $a_j>1$ and $j>\ell$: } In this
case, $\{a_1, \cdots, a_{\ell-1}\}$ are  $\ell-1$ smallest numbers
in $\{a_1, \cdots, a_{n}\}\backslash \{a_j\}.$ By induction
assumption on $t-1$, we have
$$
\begin{array}{ll}
&{a_j^{-1}}\cdot \Theta_{t-1, n-1, \ell-1}(a_1, \cdots, a_{j-1}, a_{j+1},
 \cdots, a_n)\\
 =&a_j^{-1}\cdot \Pr\big{[}\sum_{i=1}^{\ell-1}x_i\le t-1:\quad x_i\leftarrow [a_i]\big{]}
\end{array}
$$
In addition, if $a_j>a_{\ell},$ $\{a_1, \cdots, a_{\ell}\}$ are
$\ell$ smallest numbers in  $\{a_1, \cdots, a_{j-1}, a_j-1, a_{j+1},
\cdots, a_n\}$. Hence,

$$
\begin{array}{ll}
 &(1-{a_j^{-1}})\Theta_{t-1, n, \ell}(a_1, \cdots, a_{j-1}, a_j-1, a_{j+1}, \cdots, a_n)\\
 =&(1-{a_j^{-1}})\cdot \Pr\big{[}\sum_{i=1}^{\ell}x_i\le t-1:\quad x_i\leftarrow [a_i]\big{]}
\end{array}
$$

Therefore, in Eq. (\ref{eq: reduce}), we have that $Right-Left$
equals
\begin{equation}
\Pr[\sum_{i=1}^\ell x_i=t]+a_j^{-1}\cdot \Pr[\sum_{i=1}^\ell x_i\le
t-1]-a_j^{-1}\cdot \Pr[\sum_{i=1}^{\ell-1} x_i\le t-1]
\end{equation}
We need to  show  $Right-Left\ge 0.$ We split event
$\sum_{i=1}^{\ell-1}x_i\le t-1$ into two sub-events $A:
(t-1\ge)\sum_{i=1}^{\ell-1}x_i\ge t-a_\ell$ and $B:
\sum_{i=1}^{\ell-1}x_i\le t-1-a_\ell.$ Note in case of event $A$,
there exists $1\le x_\ell^*\le a_\ell$ such that
$x_\ell^*+\sum_{i=1}^{\ell-1}x_i=t.$ Hence, $\Pr[\sum_{i=1}^\ell
x_i=t]-\Pr[A]\ge \Pr[\sum_{i=1}^\ell x_i=t\wedge
x_\ell=x_\ell^*]-a_j^{-1}\Pr[A] =a_\ell^{-1}\Pr[A]-a_j^{-1}\Pr[A]\ge
0.$ In case of event $B$, since $x_\ell\le a_\ell$ always holds,
$a_j^{-1}\Pr[B]\le a_j^{-1}\Pr[\sum_{i=1}^\ell x_i\le t-1].$ Hence,
$Right\ge Left$ holds in this case.

If $a_j\le a_\ell$, then $a_j=a_\ell$ since by assumption $a_j\ge
a_\ell$ for $j>\ell$ holds always.  In this case, $\{a_1, \cdots,
a_{\ell-1}, a_\ell-1\}$ are $\ell$ smallest numbers among $\{a_1,
\cdots, a_{j-1}, a_j-1, a_{j+1}, \cdots, a_n\}.$ Hence,
$$
\begin{array}{ll}
 &(1-{a_j^{-1}})\Theta_{t-1, n, \ell}(a_1, \cdots, a_{j-1}, a_j-1, a_{j+1}, \cdots, a_n)\\
 =&(1-{a_\ell^{-1}})\cdot \Pr\big{[}x_\ell^*+\sum_{i=1}^{\ell-1}x_i\le t-1:
   x_i\leftarrow [a_i], x_\ell^*\leftarrow [a_\ell-1]\big{]}\\
 =&(1-{a_\ell^{-1}}) \sum_{u=1}^{a_{\ell}-1}\Pr\big{[}x_\ell^*+\sum_{i=1}^{\ell-1}x_i\le t-1\wedge x_\ell^*=u:
 x_i\leftarrow [a_i], x_\ell^*\leftarrow [a_\ell-1]\big{]}\\
 =&{a_\ell^{-1}} \sum_{u=1}^{a_{\ell}-1}\Pr\big{[}u+1+\sum_{i=1}^{\ell-1}x_i\le t: 
 x_i\leftarrow [a_i], i<\ell\big{]}\\
 =&\sum_{u=1}^{a_{\ell}-1}\Pr\big{[}\sum_{i=1}^{\ell}x_i\le t\wedge x_\ell=u+1:
  x_i\leftarrow [a_i], i\le\ell\big{]}\\
 =&\Pr\big{[}\sum_{i=1}^{\ell}x_i\le t\wedge x_\ell>1:\quad x_i\leftarrow [a_i]\big{]}

\end{array}
$$

Further, ${a_j^{-1}}\cdot \Theta_{t-1, n-1, \ell-1}(a_1,
\cdots, a_{j-1}, a_{j+1},
 \cdots, a_n)$

$= a_\ell^{-1}\cdot \Pr\big{[}\sum_{i=1}^{\ell-1}x_i\le t-1: \quad x_i\leftarrow [a_i]\big{]}$

 $=\Pr\big{[}\sum_{i=1}^{\ell}x_i\le t\wedge x_\ell=1:  \quad x_i\leftarrow [a_i]\big{]}.
$
    Combining  the above two equations, we have that in this case $Left=Right$.

\vspace{0.05in}\noindent{\bf Case $a_j>1$ and $j\le \ell$: } In this case, $\{a_1, \cdots, a_{\ell}\}\backslash \{a_j\}$ are $\ell-1$ smallest numbers among
$\{a_1, \cdots, a_{n}\}\backslash \{a_j\}.$ By induction assumption on $t-1$, we have
$$
\begin{array}{ll}
&{a_j^{-1}}\cdot \Theta_{t-1, n-1, \ell-1}(a_1, \cdots, a_{j-1}, a_{j+1},
 \cdots, a_n)\\
 =&a_j^{-1}\cdot \Pr\big{[}\sum_{1\le i\le \ell, i\ne j}x_i\le t-1:\quad x_i\leftarrow [a_i]\big{]}\\
 =&\Pr\big{[}\sum_{1\le i\le \ell}x_i\le t\wedge x_j=1:\quad x_i\leftarrow [a_i]\big{]}

\end{array}
$$
Note $\{a_1, \cdots, a_{j-1}, a_j-1, a_{j+1}, \cdots, a_{\ell}\}$
are the $\ell$ smallest in  $\{a_1, \cdots, a_{j-1}, a_j-1, a_{j+1},
\cdots, a_n\}$. Hence,

\hspace{-0.3in} $
\begin{array}{ll}
  &  \hspace{0.14in} (1-{a_j^{-1}})\Theta_{t-1, n, \ell}(a_1, \cdots, a_{j-1}, a_j-1, a_{j+1}, \cdots, a_n)\\
 &=(1-{a_j^{-1}}) \Pr\big{[}x_j^*+\sum_{i=1, i\ne j}^{\ell}x_i\le t-1:
 x_i\leftarrow [a_i], x_j^*\leftarrow [a_j-1]\big{]}\\
& =(1-{a_j^{-1}}) \sum_{u=1}^{a_{\ell}-1}\Pr\big{[}x_j^*+\sum_{\stackrel{i=1}{i\ne j}}^{\ell}x_i\le t-1\wedge x_j^*=u:
 x_i\leftarrow [a_i],i\ne j,  x_j^*\leftarrow [a_j-1]\big{]}\\
& ={a_j^{-1}} \sum_{u=1}^{a_{\ell}-1}\Pr\big{[}u+1+\sum_{{i=1},{i\ne j}}^{\ell}x_i\le t:
 x_i\leftarrow [a_i], i\ne j\big{]}\\
& =\sum_{u=1}^{a_{\ell}-1}\Pr\big{[}\sum_{i=1}^{\ell}x_i\le t\wedge x_j=u+1:\quad x_i\leftarrow [a_i]\big{]}\\
& =\Pr\big{[}\sum_{i=1}^{\ell}x_i\le t\wedge x_j>1:\quad x_i\leftarrow [a_i]\big{]}

\end{array}
$

Combining  the above two equations, we conclude the result in this case. 

As a summary, the induction holds for all cases. This completes the
proof. $\hfill\blacksquare$


\begin{thebibliography}{99}


\bibitem{Bao03} Feng Bao, Security Analysis of a Password Authenticated Key Exchange Protocol. {\em ISC 2003}, pages 208-217.

\bibitem{bck98} M. Bellare, R. Canetti, and H. Krawczyk, A Modular
Approach to the Design and Analysis of Authentication and Key Exchange
Protocols, {\em STOC'98}.

 \bibitem{BPR00} M. Bellare, D. Pointcheval, P. Rogaway:
Authenticated Key Exchange Secure against Dictionary Attacks.
{\em EUROCRYPT 2000}.


\bibitem{BM92} Bellovin, S.M.; Merritt, M., Encrypted key exchange:
password-based protocols secure against dictionary attacks,
In {\em Proceedings of the 1992 IEEE Computer Society
Symposium on Research in Security and Privacy,} 72-84.

\bibitem{BM93} S. M. Bellovin, M. Merritt: Augmented Encrypted
Key Exchange: A Password-Based Protocol Secure against Dictionary Attacks
and Password File Compromise. {\em ACM CCS'93}.

\bibitem{Boy99} M. K. Boyarsky, Public-key cryptography and password
protocols: the multi-user case, {\em CCS'99.}




\bibitem{ck01} R. Canetti and H. Krawczyk, Analysis of Key-Exchange
Protocols and Their Use for Building Secure Channels, {\em Eurocrypt
2001}: 453-474.





\bibitem{CBHM04} K. R. Choo, C. Boyd, Y. Hitchcock and  G. Maitland, On Session Identifiers in Provably Secure Protocols: The Bellare-Rogaway Three-Party Key Distribution Protocol Revisited. {\em SCN'04}.

\bibitem{CS01} R. Cramer and V. Shoup. Universal hash proofs and a paradigm for adaptive chosen
ciphertext secure public-key encryption. {\em
EUROCRYPT 2002}.


\bibitem{HOW92} W. Diffie, P.C. van Oorschot, and M.J. Wiener,
Authentication and Authenticated Key Exchanges, {\em Designs, Codes and
Cryptography,} vol. 2, no. 2, 1992, pp. 107-125.



\bibitem{GL03} R. Gennaro, Y. Lindell: A Framework for
Password-Based Authenticated Key Exchange. {\em EUROCRYPT 2003}: 524-543.

\bibitem{Gen08} R. Gennaro, Faster and Shorter Password-Authenticated Key Exchange, {\em TCC'08}.

 \bibitem{GL01} O. Goldreich, Y. Lindell: Session-Key Generation
Using Human Passwords Only. {\em CRYPTO'01}.

\bibitem{Gong93} L. Gong, T. Mark, A. Lomas, R. M. Needham, J. H.
Saltzer: Protecting Poorly Chosen Secrets from
     Guessing Attacks. IEEE Journal on Selected Areas in Communications
11(5): 648-656 (1993).

\bibitem{HK98} S. Halevi, H. Krawczyk, Public-Key Cryptography and
Password Protocols. {\em ACM CCS'98}.

\bibitem{HK99} S. Halevi, H. Krawczyk, Public-Key Cryptography and
Password Protocols. {\em ACM Trans. Inf. Syst. Secur.}, 2(3):
230-268, 1999.


\bibitem{HK07} D. Hofheinz and  E. Kiltz,  Secure Hybrid Encryption from Weakened
Key Encapsulation. {\em CRYPTO 2007}.

\bibitem{HK09} D. Hofheinz and  E. Kiltz, Practical Chosen Ciphertext Secure Encryption from Factoring, {\em EUROCRYPT'09}.

\bibitem{Jablon97} D. P. Jablon, Extended Password Key Exchange
Protocols
Immune to Dictionary Attacks. {\em WETICE 1997}: 248-255.

\bibitem{SAC04} S. Jiang and G. Gong, Password based Key
Exchange with Mutual Authentication, {\em SAC 2004}.



\bibitem{KOY01} J. Katz, R. Ostrovsky, M. Yung:
Efficient
Password-Authenticated Key Exchange Using Human-Memorable Passwords.
{\em EUROCRYPT'01}.



\bibitem{KV10} J. Katz and V. Vaikuntanathan, One-round Password-Based Authentication Key Exchange, {\em iacr eprint 2010/368}.

\bibitem{KV09} J. Katz and V. Vaikuntanathan, Smooth Projective Hashing and Password-Based
Authenticated Key Exchange from Lattices, {\em ASIACRYPT'09}.

\bibitem{GK10} A. Groce and J. Katz, A New Framework for Efficient
Password-Based Authenticated Key Exchange, {\em ACM CCS'10}.

\bibitem{K03} H. Krawczyk, SIGMA: The 'SIGn-and-MAc' Approach to Authenticated Diffie-Hellman and Its Use in the IKE-Protocols, {\em CRYPTO 2003}, pp. 400-425.



\bibitem{KD04} K. Kurosawa and Y. Desmedt, A New Paradigm of Hybrid Encryption
Scheme, {\em CRYPTO'04}.



\bibitem{Lucks97} S. Lucks, Open Key Exchange: How to Defeat
Dictionary
Attacks Without Encrypting Public Keys. {\em Security Protocols Workshop}
1997, pages
79-90.






\end{thebibliography}
\end{document}